\documentclass[a4paper,12pt]{article}
\usepackage{rotating}
\usepackage{graphicx,subfig}
\usepackage{latexsym}
\usepackage{amssymb}
\usepackage{amsmath}
\usepackage{bm}
\usepackage{epstopdf}
\usepackage{titlesec}
\newcommand{\squeezeup}{\vspace{-9.5mm}}

\title{Multi-place shifted nonlocal reductions of a multi-component AKNS system}

\author{Metin G\"{u}rses \thanks{gurses@fen.bilkent.edu.tr}\\
{\small Department of Mathematics, Faculty of Science}\\
{\small Bilkent University, 06800 Ankara - Turkey}\\
Asl{\i} Pekcan \thanks{Email: aslipekcan@hacettepe.edu.tr} \\
{\small Department of Mathematics, Faculty of Science} \\
{\small Hacettepe University, 06800 Ankara - Turkey}
}

\date{\nonumber}
\begin{document}
\maketitle
\date{\nonumber}
\newtheorem{thm}{Theorem}[section]
\newtheorem{Le}{Lemma}[section]
\newtheorem{defi}{Definition}[section]
\newtheorem{ex}{Example}[section]
\newtheorem{pro}{Proposition}[section]
\baselineskip 17pt

\numberwithin{equation}{section}

\begin{abstract}
Starting from a multi-component AKNS system, we obtain new shifted nonlocal nonlinear Schr\"{o}dinger equations. We find 13 different shifted nonlocal nonlinear Schr\"{o}dinger equations with two-place nonlocalities and 10 shifted nonlocal nonlinear Schr\"{o}dinger equations with four-place nonlocalities. We first obtain one-soliton solutions of the multi-component AKNS system by the Hirota method. Applying the shifted nonlocal reduction formulas to this solution, we obtain one-soliton solutions for the shifted nonlocal nonlinear Schr\"{o}dinger equations. In cases yielding nontrivial solutions, we discuss the singularity structures of the solutions and show that the one-soliton solutions we obtain are nonsingular for certain values of the parameters. We plot representative nonsingular solutions obtained for admissible parameter values.
\end{abstract}


\section{Introduction}

A multi-component Ablowitz-Kaup-Newell-Segur (AKNS) system \cite{MA2019} is
\begin{align}
&ip_{j,t}+p_{j,xx}+2\sigma (\sum_{l=1}^N p_lq_l)p_j=0,\quad j=1, \ldots, N, \label{generalmultia}\\
&iq_{j,t}-q_{j,xx}-2\sigma (\sum_{l=1}^N p_lq_l)q_j=0,\quad j=1, \ldots, N, \label{generalmultib}
\end{align}
where $p_j=p_j(x,t)$, $q_j=q_j(x,t)$, $j=1, \ldots, N$ are complex dynamical variables and $\sigma=\pm 1$.
The recursion operator of the above system is $\mathcal{R}=J_2J_1^{-1}$
where $(J_1, J_2)$ is the Hamiltonian pair
given by
{\small
\begin{align}\displaystyle
&J_1=\left( \begin{array}{cc}
0 &I_N \\
-I_N& 0
 \end{array} \right),\\
 &\nonumber\\
&J_2=\left( \begin{array}{cc}
\sigma(p^T\partial^{-1}p+(p^T\partial^{-1}p)^T) &-(\partial +\sigma \sum_{j=1}^N p_j\partial^{-1}q_j)I_N-\sigma p^T\partial^{-1}q^T \\
-(\partial +\sigma \sum_{j=1}^N q_j\partial^{-1}p_j)I_N-\sigma q\partial^{-1}p& \sigma (q\partial^{-1}q^T+(q\partial^{-1}q^T)^T)
 \end{array} \right).
\end{align}}
\noindent Here $\partial^{-1}$ denotes the standard antiderivative with respect to $x$, $I_N$ is the $N\times N$ identity matrix for $N\geq 2$, $p=(p_1,\ldots, p_N)$, $q=(q_1,\ldots, q_N)^T$,
and $T$ denotes matrix transpose. We have the multi-component hierarchy of AKNS
\begin{equation}
iU_t=\left(\mathcal{R}_N \right)^n U_x, ~~~~n=1, 2, \ldots,\quad U=(p,q^T)^T,
\end{equation}
where
\begin{equation}
\mathcal{R}_{N}=\left( \begin{array}{cc}
R_{1} & R_{2} \\
R_{3}& R_{4}
\end{array} \right),
\end{equation}
with
\begin{eqnarray}
&&\left(R_{1}\right)^{i}\,_{j}=-\left(\partial+\sigma p^{k}\, \partial^{-1}\, q_{k} \right)\, \delta^{i}\,_{j}-\sigma\, p^{i}\, \partial^{-1}\, q_{j}, \\
&&\left(R_{2}\right)^{i}\,_{j}=-\sigma\, \left(p^{i} \partial^{-1}\, p^{j} +p^{j} \, \partial^{-1}\, p^{i}\,\right), \\
&&\left(R_{3}\right)^{i}\,_{j}=\sigma\, \left(q^{i} \partial^{-1}\, q_{j} +q_{j} \, \partial^{-1}\, q^{i}\,\right), \\
&&\left(R_{4}\right)^{i}\,_{j}= \left(\partial+\sigma q_{k}\, \partial^{-1}\, p^{k} \right)\, \delta^{i}\,_{j}+\sigma\, q^{i}\, \partial^{-1}\, p^{j},
\end{eqnarray}
for $i, j=1, 2, \cdots, N$. Then the evolution equations are obtained from
\begin{eqnarray}
&&ip^{m}_{t}=\left(R_{1}\right)^{m}\,_{j}\,p^{j}_{x}+\left(R_{2}\right)^{m}\,_{j}\,q^{j}_{x}, ~~~m=1,2, \cdots, N,\\
&&iq^{m}_{t}=\left(R_{3}\right)^{m}\,_{j}\,p^{j}_{x}+\left(R_{4}\right)^{m}\,_{j}\,q^{j}_{x}, ~~~m=1,2, \cdots, N.
\end{eqnarray}
Here we use the Einstein convention where repeated indices are summed up from $1$ to $N$.

\noindent For $N=2$, we have the recursion operator $\mathcal{R}_2$ given explicitly as
\begin{equation}
\mathcal{R}_2=\left( \begin{array}{cccc}
A_{11}& A_{12} & A_{13} & A_{14} \\
A_{21}& A_{22} & A_{23} & A_{24} \\
A_{31}& A_{32} & A_{33} & A_{34} \\
A_{41}& A_{42} & A_{43} & A_{44}
 \end{array} \right),
\end{equation}
where
\begin{align}
&A_{11}=-(\partial+\sigma(2p_1\partial^{-1}q_1+p_2\partial^{-1}q_2)),\quad A_{12}=-\sigma p_1\partial^{-1}q_2,\\
&A_{13}=-2\sigma p_1\partial^{-1}p_1,\quad A_{14}=-\sigma (p_1\partial^{-1}p_2+p_2\partial^{-1}p_1),\\
&A_{21}=-\sigma p_2\partial^{-1} q_1,\quad A_{22}=-(\partial+\sigma(p_1\partial^{-1}q_1+2p_2\partial^{-1} q_2)),\\
&A_{23}= -\sigma(p_2\partial^{-1}p_1+p_1\partial^{-1}p_2),\quad A_{24}=  -2\sigma p_2\partial^{-1} p_2,\\
&A_{31}=2\sigma q_1\partial^{-1}q_1,\quad A_{32}= \sigma(q_1\partial^{-1}q_2+q_2\partial^{-1}q_1),\\
&A_{33}= (\partial+\sigma(2q_1\partial^{-1}p_1+q_2\partial^{-1}p_2)),\quad A_{34}=\sigma q_1\partial^{-1} p_2,\\
&A_{41}=\sigma (q_2\partial^{-1}q_1+q_1\partial^{-1}q_2),\quad A_{42}=2\sigma q_2\partial^{-1}q_2,\\
&A_{43}= \sigma q_2\partial^{-1} p_1,\quad A_{44}=(\partial +\sigma (q_1\partial^{-1}p_1+2q_2\partial^{-1}p_2)).
\end{align}
Hence the hierarchy obtained here is
\begin{equation}
i\left( \begin{array}{c}
p_1 \\
p_2\\
q_1\\
q_2
 \end{array} \right)_t=\mathcal{R}_2^m \left( \begin{array}{c}
p_1 \\
p_2\\
q_1\\
q_2
 \end{array} \right)_x,\quad m=1,2, \ldots\, .
\end{equation}

Let $m=1$, $p_1=q, p_2=p, q_1=r$, and $q_2=s$. In order to obtain a richer family of shifted nonlocal equations we let also $t\rightarrow \frac{c}{i} t$ so that the fixed constant $i$ becomes an arbitrary parameter $c$. Hence we get the following four-component AKNS system \cite{MA2002}
\begin{align}\displaystyle
&cq_t+q_{xx}+2\sigma [qr+ps]q=0,\label{4NLS-a}\\
&cp_t+p_{xx}+2\sigma [qr+ps]p=0,\label{4NLS-b}\\
&cr_t-r_{xx}-2\sigma [qr+ps]r=0,\label{4NLS-c}\\
&cs_t-s_{xx}-2\sigma [qr+ps]s=0.\label{4NLS-d}
\end{align}
 Indeed in \cite{MA2002}, Ma and Zhou constructed a multi-component
AKNS hierarchy and the above system is the first member of this hierarchy. They generated the Hamiltonian structure of this multi-component hierarchy.
In \cite{YWL}, the authors derived a more general multi-component AKNS hierarchy including the one obtained in \cite{MA2002}.

Physical models are usually defined in terms of a local space-time point $(x,t)$, where $x$ is the spatial coordinate and $t$ is the time. However, in the natural and social sciences, we frequently encounter events happening at different points in space and time related to one another. Modeling such related events, which happen at multi-places, i.e., at different locations and different times, is usually achieved through models different from the classical local models. Nonlocal nonlinear partial differential equations are suitable for modeling such phenomena. This idea was first introduced by Ablowitz and Musslimani \cite{abl1}. Unlike the local models, the nonlocal models contain, in addition to the terms evaluated at the point $(x,t)$, terms from the reflected space and time points such as $(-x,t)$ (space reversal), $(x,-t)$ (time reversal), and $(-x,-t)$ (space-time reversal). An important property of the nonlocal reductions is the fact that when consistently applied, the reductions preserve the integrability of the original model. Nonlocal reductions are therefore a tool for constructing new integrable models from already known integrable ones.

In a series of papers, Ablowitz and Musslimani studied $(1+1)$-dimensional unshifted nonlocal equations, including nonlocal nonlinear Schr\"{o}dinger (NLS), nonlocal modified Korteweg-de Vries (mKdV), and nonlocal sine-Gordon equations \cite{abl1}, \cite{abl2} using the inverse scattering transform (IST) method and Riemann-Hilbert approach. In \cite{abl3}, they further studied the $(2+1)$-dimensional Davey-Stewartson (DS) system and three-wave interaction system in their nonlocal unshifted forms. The two-place shifted nonlocal NLS equation and its soliton solutions are studied in \cite{AbMu4}. Later, Ablowitz et al. considered two-place shifted NLS equations with time reversal, space reversal, and space-time reversal nonlocalities, as well as their integrable discretization, and derived one-soliton solutions using IST method \cite{AbMu5}. Subsequent works by G\"{u}rses and Pekcan investigated a number of different unshifted and shifted nonlocal reductions of classical integrable equations, such as NLS, mKdV, KdV type equations, and also a number of $(2+1)$-dimensional nonlocal partial differential equations \cite{gur1}-\cite{gur6}. All the nonlocal single equations obtained in these works are of the two-place type. In \cite{GPZ}, it was proved that unshifted nonlocal reductions can be interpreted as special cases of unshifted discrete symmetry transformations, which are a special type of unshifted scale transformations. A similar result was given for shifted nonlocal reductions in \cite{Bayli}.

In recent years, the study of multi-place nonlocal equations has drawn significant attention. These equations have been extensively investigated by Lou and his collaborators \cite{Loumulti}-\cite{SYLou2}, as well as by Ma in \cite{Mamulti}. In particular, Lou has obtained unshifted two- and four-place nonlocal Kadomtsev–Petviashvili (KP) and NLS equations in \cite{Loumulti}, and shifted multi-place nonlocal KdV equations in \cite{SYLou1}. In the recent paper \cite{Mamulti}, Ma has derived unshifted multi-place nonlocal NLS equations. In \cite{Bayli}, Bayl{\i} and Pekcan have presented several integrable two-place and four-place shifted nonlocal Maccari equations, and their one-soliton solutions by the Hirota method. There have not been many studies conducted on the topic of four-place shifted nonlocal equations in the literature.

To find solutions of the reduced nonlocal systems, we use our previously introduced method \cite{gur1}, \cite{gur3}. This method is based on using a solution of the original system with the nonlocal reduction formulas.

\noindent
(a) Consider a system with dynamical variables $r(x,t)$ and $q(x,t)$ with solutions

$$r(x,t)=\frac{W_1(x,t)}{W_2(x,t)},~~~q(x,t)=\frac{V_1(x,t)}{V_2(x,t)}.$$

(b) Let us now consider the shifted nonlocal reduction formula $r(x,t)=q(\varepsilon_1x+x_0,\varepsilon_2t+t_0)$ where $\varepsilon_1^2=\varepsilon_2^2=1$, $x_0, t_0 \in \mathbb{R}$.

\noindent
(c) The reduction formula gives the following relations for the functions $W_1, W_2, V_1,$ and $V_2$:
\begin{equation}\label{types}
\frac{W_1(x,t)}{W_2(x,t)}= \frac{V_1(\varepsilon_1x+x_0,\varepsilon_2t+t_0)}{V_2(\varepsilon_1x+x_0,\varepsilon_2t+t_0)}.
\end{equation}

\noindent
(d) As a sufficient reduction ansatz, we impose the equality of the numerators and denominators of $r(x,t)$ and $q(x,t)$ separately. Hence from \eqref{types} we have
\begin{equation}\label{type1}
W_1(x,t)=V_1(\varepsilon_1x+x_0,\varepsilon_2t+t_0),\quad W_2(x,t)=V_2(\varepsilon_1x+x_0,\varepsilon_2t+t_0).
\end{equation}

\noindent
(e) From the equations \eqref{type1} we obtain constraints on the parameters of the solutions. Using these constraints on the solution $q(x,t)$ gives
the solution of the shifted nonlocal equation reduced from the system by $r(x,t)=q(\varepsilon_1x+x_0,\varepsilon_2t+t_0)$.

\noindent
In this work we show that the four-component AKNS system (\ref{4NLS-a})-(\ref{4NLS-d}) reduces to
two- and four-place shifted nonlocal nonlinear Schr\"{o}dinger (NLS) equations. In Section 2, we give all possible two-place and four-place shifted nonlocal equations. In Section 3,
we use the Hirota bilinear method to obtain soliton solutions of the four-component AKNS system (\ref{4NLS-a})-(\ref{4NLS-d}). Using this solution with the
shifted nonlocal reduction formulas via our method, we obtain one-soliton solutions of two-place and four-place shifted nonlocal NLS equations in Sections 4 and 5, respectively.

\section{Shifted nonlocal reduced equations}

Applying shifted nonlocal reductions to the four-component AKNS system (\ref{4NLS-a})-(\ref{4NLS-d}) consistently, we derive
two-place and four-place shifted nonlocal NLS equations.

Let us present how we use the shifted nonlocal reductions on the system (\ref{4NLS-a})-(\ref{4NLS-d}) with an example. We first use
the reductions
\begin{equation}
p(x,t)=q(\varepsilon_1x+x_0,\varepsilon_2t+t_0),\quad r(x,t)=s(\varepsilon_1x+x_0,\varepsilon_2t+t_0),
\end{equation}
where $\varepsilon_i^2=1$, $i=1, 2$, and $x_0, t_0 \in \mathbb{R}$. Under these reductions (\ref{4NLS-a}) becomes
\begin{equation}\label{q-red}
cq_t(x,t)+q_{xx}(x,t)+2\sigma[q(x,t)s(\varepsilon_1x+x_0,\varepsilon_2t+t_0)+q(\varepsilon_1x+x_0,\varepsilon_2t+t_0)s(x,t)]q(x,t)=0,
\end{equation}
and (\ref{4NLS-b}) takes the form
\begin{align}
&cq_t(\varepsilon_1x+x_0,\varepsilon_2t+t_0)+q_{xx}(\varepsilon_1x+x_0,\varepsilon_2t+t_0)
\nonumber\\&+2\sigma[q(x,t)s(\varepsilon_1x+x_0,\varepsilon_2t+t_0)+q(\varepsilon_1x+x_0,\varepsilon_2t+t_0)s(x,t)]q(\varepsilon_1x+x_0,\varepsilon_2t+t_0)=0.
\end{align}
Letting $\tilde{x}=\varepsilon_1x+x_0$ and $\tilde{t}=\varepsilon_2t+t_0$, the above equation becomes
\begin{align}\label{p-red}
&c\varepsilon_2q_{\tilde{t}}(\tilde{x},\tilde{t})+q_{\tilde{x}\tilde{x}}(\tilde{x},\tilde{t})
\nonumber\\&+2\sigma[q(\varepsilon_1(\tilde{x}-x_0),\varepsilon_2(\tilde{t}-t_0))
s(\tilde{x},\tilde{t})+q(\tilde{x},\tilde{t})s(\varepsilon_1(\tilde{x}-x_0),\varepsilon_2(\tilde{t}-t_0))]q(\tilde{x},\tilde{t})=0.
\end{align}
Hence to have the equations (\ref{q-red}) and (\ref{p-red}) equivalent we must have $\varepsilon_2=1$, $t_0=0$, and $x_0=0$ if $\varepsilon_1=1$, and
$x_0=\mathrm{arbitrary}$ if $\varepsilon_1=-1$. Similarly, from (\ref{4NLS-c}) and (\ref{4NLS-d}) we get the same conditions. To obtain a nonlocal reduction we choose $\varepsilon_1=-1$. Hence the four-component AKNS system (\ref{4NLS-a})-(\ref{4NLS-d}) consistently reduces to
\begin{align}
cq_t(x,t)+q_{xx}(x,t)+2\sigma[q(x,t)s(-x+x_0,t)+q(-x+x_0,t)s(x,t)]q(x,t)=0,\label{q-one}\\
cs_t(x,t)-s_{xx}(x,t)-2\sigma[q(x,t)s(-x+x_0,t)+q(-x+x_0,t)s(x,t)]s(x,t)=0.\label{s-one}
\end{align}
If we apply shifted nonlocal reduction one more time to the above system, e.g. $s(x,t)=\bar{q}(\mu_1x+X_0,\mu_2t+T_0)$, $\mu_i^2=1$, $i=1, 2$, $X_0, T_0 \in \mathbb{R}$ we get
the following constraints for consistency
\begin{equation}
c=-\bar{c}\mu_2;\quad  T_0=0\quad \mathrm{if}\quad  \mu_2=1;\quad  X_0=0\quad  \mathrm{if}\quad  \mu_1=1;\quad  X_0=x_0\quad  \mathrm{if}\quad  \mu_1=-1.
\end{equation}
If we choose $(\mu_1,\mu_2)=(-1,1)$ then $c=-\bar{c}$ and the system (\ref{q-one})-(\ref{s-one}) reduces to the following single equation
\begin{equation*}
cq_t(x,t)+q_{xx}(x,t)+2\sigma[q(x,t)\bar{q}(x,t)+q(-x+x_0,t)\bar{q}(-x+x_0,t)]q(x,t)=0,
\end{equation*}
consistently. Note that this equation is the first equation given in Section 2.1.

In the same manner, we systematically apply the admissible shifted nonlocal reductions on the four-component AKNS system (\ref{4NLS-a})-(\ref{4NLS-d}) and obtain all possible two-place and four-place shifted nonlocal NLS equations.

 \subsection{Two-place nonlocal reduced equations}

We have 13 different two-place shifted nonlocal reductions of the four-component AKNS system (\ref{4NLS-a})-(\ref{4NLS-d}).

\noindent 1)\, $p(x,t)=q(-x+x_0,t)$, $r(x,t)=\bar{q}(x,t)$, $s(x,t)=\bar{q}(-x+x_0,t)$, $c=-\bar{c}$.
\begin{equation}\label{twopex2}
cq_t(x,t)+q_{xx}(x,t)+2\sigma[q(x,t)\bar{q}(x,t)+q(-x+x_0,t)\bar{q}(-x+x_0,t)]q(x,t)=0.
\end{equation}

\noindent 2)\, $p(x,t)=q(-x+x_0,t)$, $r(x,t)=\bar{q}(-x+x_0,t)$, $s(x,t)=\bar{q}(x,t)$, $c=-\bar{c}$.
\begin{equation}\label{twopex3}
cq_t(x,t)+q_{xx}(x,t)+2\sigma[q(x,t)\bar{q}(-x+x_0,t)+q(-x+x_0,t)\bar{q}(x,t)]q(x,t)=0.
\end{equation}

\noindent 3)\, $p(x,t)=q(x,t)$, $r(x,t)=q(x,-t+t_0)$, $s(x,t)=q(x,-t+t_0)$.
\begin{equation}\label{twopex4}
cq_t(x,t)+q_{xx}(x,t)+4\sigma q(x,t)^2q(x,-t+t_0)=0.
\end{equation}

\noindent 4)\, $p(x,t)=q(x,t)$, $r(x,t)=q(-x+x_0,-t+t_0)$, $s(x,t)=q(-x+x_0,-t+t_0)$.
\begin{equation}\label{twopex5}
cq_t(x,t)+q_{xx}(x,t)+4\sigma q(x,t)^2q(-x+x_0,-t+t_0)=0.
\end{equation}

\noindent 5)\, $p(x,t)=q(x,t)$, $r(x,t)=\bar{q}(x,-t+t_0)$, $s(x,t)=\bar{q}(x,-t+t_0)$, $c=\bar{c}$.
\begin{equation}\label{twopex6}
cq_t(x,t)+q_{xx}(x,t)+4\sigma q(x,t)^2\bar{q}(x,-t+t_0)=0.
\end{equation}

\noindent 6)\, $p(x,t)=q(x,t)$, $r(x,t)=\bar{q}(-x+x_0,t)$, $s(x,t)=\bar{q}(-x+x_0,t)$, $c=-\bar{c}$.
\begin{equation}\label{twopex7}
cq_t(x,t)+q_{xx}(x,t)+4\sigma q(x,t)^2\bar{q}(-x+x_0,t)=0.
\end{equation}

\noindent 7)\, $p(x,t)=q(x,t)$, $r(x,t)=\bar{q}(-x+x_0,-t+t_0)$, $s(x,t)=\bar{q}(-x+x_0,-t+t_0)$, $c=\bar{c}$.
\begin{equation}\label{twopex8}
cq_t(x,t)+q_{xx}(x,t)+4\sigma q(x,t)^2\bar{q}(-x+x_0,-t+t_0)=0.
\end{equation}

\noindent 8)\, The following two reductions lead to the same equation: Case i. $p(x,t)=\bar{q}(x,t)$, $r(x,t)=q(x,-t+t_0)$, $s(x,t)=\bar{q}(x,-t+t_0)$, $c=\bar{c}$, Case ii. $p(x,t)=\bar{q}(x,-t+t_0)$, $r(x,t)=q(x,-t+t_0)$, $s(x,t)=\bar{q}(x,t)$, $c=-\bar{c}$.
\begin{equation}\label{twopex9}
cq_t(x,t)+q_{xx}(x,t)+2\sigma[q(x,t)q(x,-t+t_0)+\bar{q}(x,-t+t_0)\bar{q}(x,t)]q(x,t)=0.
\end{equation}

\noindent 9)\, The following two reductions lead to the same equation: Case i. $p(x,t)=\bar{q}(-x+x_0,-t+t_0)$, $r(x,t)=\bar{q}(x,t)$, $s(x,t)=q(-x+x_0,-t+t_0)$, $c=-\bar{c}$, Case ii. $p(x,t)=\bar{q}(-x+x_0,-t+t_0)$, $r(x,t)=q(-x+x_0,-t+t_0)$, $s(x,t)=\bar{q}(x,t)$, $c=-\bar{c}$.
\begin{equation}\label{twopex10}
cq_t(x,t)+q_{xx}(x,t)+2\sigma[q(x,t)\bar{q}(x,t)+q(-x+x_0,-t+t_0)\bar{q}(-x+x_0,-t+t_0)]q(x,t)=0.
\end{equation}

\noindent 10)\, $p(x,t)=\bar{q}(x,-t+t_0)$, $r(x,t)=\bar{q}(x,t)$, $s(x,t)=q(x,-t+t_0)$, $c=-\bar{c}$.
\begin{equation}\label{twopex1}
cq_t(x,t)+q_{xx}(x,t)+2\sigma[q(x,t)\bar{q}(x,t)+q(x,-t+t_0)\bar{q}(x,-t+t_0)]q(x,t)=0.
\end{equation}

\noindent 11)\, The following two reductions lead to the same equation: Case i. $p(x,t)=\bar{q}(x,t)$, $r(x,t)=q(-x+x_0,-t+t_0)$, $s(x,t)=\bar{q}(-x+x_0,-t+t_0)$, $c=\bar{c}$, Case ii. $p(x,t)=\bar{q}(-x+x_0,-t+t_0)$, $r(x,t)=q(-x+x_0,-t+t_0)$, $s(x,t)=\bar{q}(x,t)$, $c=-\bar{c}$.
\begin{equation}\label{twopex11}
cq_t(x,t)+q_{xx}(x,t)+2\sigma[q(x,t)q(-x+x_0,-t+t_0)+\bar{q}(x,t)\bar{q}(-x+x_0,-t+t_0)]q(x,t)=0.
\end{equation}

\noindent 12)\, $p(x,t)=\bar{q}(x,t)$, $r(x,t)=\bar{q}(-x+x_0,-t+t_0)$, $s(x,t)=q(-x+x_0,-t+t_0)$, $c=\bar{c}$.
\begin{equation}\label{twopex12}
cq_t(x,t)+q_{xx}(x,t)+2\sigma[q(x,t)\bar{q}(-x+x_0,-t+t_0)+\bar{q}(x,t)q(-x+x_0,-t+t_0)]q(x,t)=0.
\end{equation}

\noindent 13)\, $p(x,t)=\bar{q}(x,t)$, $r(x,t)=\bar{q}(x,-t+t_0)$, $s(x,t)=q(x,-t+t_0)$, $c=\bar{c}$.
\begin{equation}\label{twopex13}
cq_t(x,t)+q_{xx}(x,t)+2\sigma[q(x,t)\bar{q}(x,-t+t_0)+\bar{q}(x,t)q(x,-t+t_0)]q(x,t)=0.
\end{equation}

 \subsection{Four-place nonlocal reduced equations}

We have 10 different four-place shifted nonlocal reductions of the four-component AKNS system (\ref{4NLS-a})-(\ref{4NLS-d}).

\noindent 1)\, $p(x,t)=q(-x+x_0,t)$, $r(x,t)=q(-x+x_0,-t+t_0)$, $s(x,t)=q(x,-t+t_0)$.
\begin{equation}\label{four-ex1}
cq_t(x,t)+q_{xx}(x,t)+2\sigma[q(x,t)q(-x+x_0,-t+t_0)+q(-x+x_0,t)q(x,-t+t_0)]q(x,t)=0.
\end{equation}

\noindent 2)\, $p(x,t)=q(-x+x_0,t)$, $r(x,t)=q(x,-t+t_0)$, $s(x,t)=q(-x+x_0,-t+t_0)$.
\begin{equation}\label{four-ex2}
cq_t(x,t)+q_{xx}(x,t)+2\sigma[q(x,t)q(x,-t+t_0)+q(-x+x_0,t)q(-x+x_0,-t+t_0)]q(x,t)=0.
\end{equation}

\noindent 3)\, $p(x,t)=q(-x+x_0,t)$, $r(x,t)=\bar{q}(-x+x_0,-t+t_0)$, $s(x,t)=\bar{q}(x,-t+t_0)$, $c=\bar{c}$.
\begin{equation}\label{four-ex3}
cq_t(x,t)+q_{xx}(x,t)+2\sigma[q(x,t)\bar{q}(-x+x_0,-t+t_0)+q(-x+x_0,t)\bar{q}(x,-t+t_0)]q(x,t)=0.
\end{equation}

\noindent 4)\, $p(x,t)=q(-x+x_0,t)$, $r(x,t)=\bar{q}(x,-t+t_0)$, $s(x,t)=\bar{q}(-x+x_0,-t+t_0)$, $c=\bar{c}$.
\begin{equation}\label{four-ex4}
cq_t(x,t)+q_{xx}(x,t)+2\sigma[q(x,t)\bar{q}(x,-t+t_0)+q(-x+x_0,t)\bar{q}(-x+x_0,-t+t_0)]q(x,t)=0.
\end{equation}

\noindent 5)\, $p(x,t)=\bar{q}(-x+x_0,-t+t_0)$, $r(x,t)=q(x,-t+t_0)$, $s(x,t)=\bar{q}(-x+x_0,t)$, $c=-\bar{c}$.
\begin{equation}\label{four-ex5}
cq_t(x,t)+q_{xx}(x,t)+2\sigma[q(x,t)q(x,-t+t_0)+\bar{q}(-x+x_0,-t+t_0)\bar{q}(-x+x_0,t)]q(x,t)=0.
\end{equation}

\noindent 6)\, The following two reductions lead to the same equation: Case i. $p(x,t)=\bar{q}(-x+x_0,t)$, $r(x,t)=q(-x+x_0,-t+t_0)$, $s(x,t)=\bar{q}(x,-t+t_0)$, $c=\bar{c}$, Case ii. $p(x,t)=\bar{q}(x,-t+t_0)$, $r(x,t)=q(-x+x_0,-t+t_0)$, $s(x,t)=\bar{q}(-x+x_0,t)$, $c=-\bar{c}$.
\begin{equation}\label{four-ex6}
cq_t(x,t)+q_{xx}(x,t)+2\sigma[q(x,t)q(-x+x_0,-t+t_0)+\bar{q}(-x+x_0,t)\bar{q}(x,-t+t_0)]q(x,t)=0.
\end{equation}

\noindent 7)\, $p(x,t)=\bar{q}(x,-t+t_0)$, $r(x,t)=\bar{q}(-x+x_0,t)$, $s(x,t)=q(-x+x_0,-t+t_0)$, $c=-\bar{c}$.
\begin{equation}\label{four-ex7}
cq_t(x,t)+q_{xx}(x,t)+2\sigma[q(x,t)\bar{q}(-x+x_0,t)+\bar{q}(x,-t+t_0)q(-x+x_0,-t+t_0)]q(x,t)=0.
\end{equation}

\noindent 8)\, $p(x,t)=\bar{q}(-x+x_0,t)$, $r(x,t)=\bar{q}(-x+x_0,-t+t_0)$, $s(x,t)=q(x,-t+t_0)$, $c=\bar{c}$.
\begin{equation}\label{four-ex8}
cq_t(x,t)+q_{xx}(x,t)+2\sigma[q(x,t)\bar{q}(-x+x_0,-t+t_0)+\bar{q}(-x+x_0,t)q(x,-t+t_0)]q(x,t)=0.
\end{equation}

\noindent 9)\, $p(x,t)=\bar{q}(-x+x_0,t)$, $r(x,t)=\bar{q}(x,-t+t_0)$, $s(x,t)=q(-x+x_0,-t+t_0)$, $c=\bar{c}$.
\begin{equation}\label{four-ex9}
cq_t(x,t)+q_{xx}(x,t)+2\sigma[q(x,t)\bar{q}(x,-t+t_0)+\bar{q}(-x+x_0,t)q(-x+x_0,-t+t_0)]q(x,t)=0.
\end{equation}

\noindent 10)\, $p(x,t)=\bar{q}(-x+x_0,-t+t_0)$, $r(x,t)=\bar{q}(-x+x_0,t)$, $s(x,t)=q(x,-t+t_0)$, $c=-\bar{c}$.
\begin{equation}\label{four-ex10}
cq_t(x,t)+q_{xx}(x,t)+2\sigma[q(x,t)\bar{q}(-x+x_0,t)+\bar{q}(-x+x_0,-t+t_0)q(x,-t+t_0)]q(x,t)=0.
\end{equation}

\section{The Hirota method for the multi-component AKNS}

Let $\displaystyle p_j=\frac{g_j}{f}$ and $\displaystyle q_j=\frac{h_j}{f}$, for $j=1,\ldots, N$ in the multi-component AKNS system (\ref{generalmultia}) and
(\ref{generalmultib}). We get the Hirota bilinear form of this system as
\begin{align}
&(iD_t+D_x^2)\{g_j\cdot f\}=0,\quad j=1,\ldots, N\\
&(iD_t-D_x^2)\{h_j\cdot f\}=0,\quad j=1,\ldots, N\\
&D_x^2\{f\cdot f\}=2\sigma \sum_{l=1}^N g_lh_l.
\end{align}
For the four-component AKNS system (\ref{4NLS-a})-(\ref{4NLS-d}), $N=2$ above and recall that we take $t\rightarrow \frac{c}{i} t$.

\subsection{One-soliton solution of the four-component AKNS}

Let $\displaystyle q(x,t)=\frac{g(x,t)}{f(x,t)}$, $\displaystyle r(x,t)=\frac{h(x,t)}{f(x,t)}$, $\displaystyle p(x,t)=\frac{a(x,t)}{f(x,t)}$, and $\displaystyle
s(x,t)=\frac{b(x,t)}{f(x,t)}$ in
the system (\ref{4NLS-a})-(\ref{4NLS-d}). Then we obtain the Hirota bilinear form of this system as
\begin{align}\displaystyle
&(cD_t+D_x^2)\{g\cdot f\}=0,\label{H-a}\\
&(cD_t+D_x^2)\{a\cdot f\}=0,\label{H-b}\\
&(cD_t-D_x^2)\{h\cdot f\}=0,\label{H-c}\\
&(cD_t-D_x^2)\{b\cdot f\}=0,\label{H-d}\\
&D_x^2\{f\cdot f\}=2\sigma (gh+ab).\label{H-e}
\end{align}

Use the following expansion in the Hirota bilinear form (\ref{H-a})-(\ref{H-e});
\begin{align}\displaystyle
&g=\varepsilon g_1+\varepsilon^3 g_3,\quad a=\varepsilon a_1+\varepsilon^3 a_3,\quad h=\varepsilon h_1+\varepsilon^3 h_3,\\
&b=\varepsilon b_1+\varepsilon^3 b_3,\quad f=1+\varepsilon^2 f_2+\varepsilon^4 f_4,
\end{align}
where
\begin{equation}\displaystyle
g_1=e^{\theta_1},\quad a_1=e^{\theta_2},\quad h_1=e^{\theta_3},\quad b_1=e^{\theta_4},
\end{equation}
for $\theta_j=k_jx+\omega_j t+\delta_j$, $j=1,2,3,4$. We equate the coefficients of $\varepsilon^m$, $m=1, 2, \cdots, 8$ to zero. The coefficients of
$\varepsilon$ give the dispersion relations
\begin{equation}\label{dispersion}\displaystyle
\omega_1=-\frac{1}{c}k_1^2,\quad \omega_2=-\frac{1}{c}k_2^2,\quad \omega_3=\frac{1}{c}k_3^2,\quad \omega_4=\frac{1}{c}k_4^2.
\end{equation}
From the coefficient of $\varepsilon^2$ we get
\begin{equation}\displaystyle
f_2=\sigma \Big[\frac{e^{\theta_1+\theta_3}}{(k_1+k_3)^2}+\frac{e^{\theta_2+\theta_4}}{(k_2+k_4)^2}\Big].
\end{equation}
The coefficients of $\varepsilon^3$ give
\begin{align}
&g_3=\gamma_1 e^{\theta_1+\theta_2+\theta_4},\quad a_3=\gamma_2 e^{\theta_1+\theta_2+\theta_3},\\
&h_3=\gamma_3 e^{\theta_2+\theta_3+\theta_4},\quad b_3=\gamma_4 e^{\theta_1+\theta_3+\theta_4},
\end{align}
where
\begin{align}\label{gammalar}\displaystyle
&\gamma_1=\frac{\sigma(k_1-k_2)}{(k_1+k_4)(k_2+k_4)^2}, \quad \gamma_2=-\frac{\sigma(k_1-k_2)}{(k_2+k_3)(k_1+k_3)^2},\\
&\gamma_3=\frac{\sigma(k_3-k_4)}{(k_2+k_3)(k_2+k_4)^2}, \quad \gamma_4=-\frac{\sigma(k_3-k_4)}{(k_1+k_4)(k_1+k_3)^2}.
\end{align}
From the coefficient of $\varepsilon^4$ we obtain the function $f_4$ as
\begin{equation}
f_4=Me^{\theta_1+\theta_2+\theta_3+\theta_4},
\end{equation}
where
\begin{equation}\label{M}\displaystyle
M=\frac{(k_1-k_2)(k_3-k_4)}{(k_1+k_4)(k_2+k_3)(k_1+k_3)^2(k_2+k_4)^2}.
\end{equation}
The coefficients of $\varepsilon^5$, $\varepsilon^6$, $\varepsilon^7$, and $\varepsilon^8$ vanish identically. Without loss of generality take $\varepsilon=1$.
Hence one-soliton solution of the four-component AKNS system (\ref{4NLS-a})-(\ref{4NLS-d}) is given by $(q,p,r,s)$ where
\begin{align}\displaystyle
&q=\frac{g}{f}=\frac{e^{\theta_1}+\gamma_1e^{\theta_1+\theta_2+\theta_4}}{1+\sigma\Big[\frac{e^{\theta_1+\theta_3}}{(k_1+k_3)^2}+\frac{e^{\theta_2+\theta_4}}{(k_2+k_4)^2}
\Big]+Me^{\theta_1+\theta_2+\theta_3+\theta_4}},\label{q-sol}\\
&p=\frac{a}{f}=\frac{e^{\theta_2}+\gamma_2e^{\theta_1+\theta_2+\theta_3}}{1+\sigma\Big[\frac{e^{\theta_1+\theta_3}}{(k_1+k_3)^2}+\frac{e^{\theta_2+\theta_4}}{(k_2+k_4)^2}
\Big]+Me^{\theta_1+\theta_2+\theta_3+\theta_4}},\label{p-sol}\\
&r=\frac{h}{f}=\frac{e^{\theta_3}+\gamma_3e^{\theta_2+\theta_3+\theta_4}}{1+\sigma\Big[\frac{e^{\theta_1+\theta_3}}{(k_1+k_3)^2}+\frac{e^{\theta_2+\theta_4}}{(k_2+k_4)^2}
\Big]+Me^{\theta_1+\theta_2+\theta_3+\theta_4}},\label{r-sol}\\
&s=\frac{b}{f}=\frac{e^{\theta_4}+\gamma_4e^{\theta_1+\theta_3+\theta_4}}{1+\sigma\Big[\frac{e^{\theta_1+\theta_3}}{(k_1+k_3)^2}+\frac{e^{\theta_2+\theta_4}}{(k_2+k_4)^2}
\Big]+Me^{\theta_1+\theta_2+\theta_3+\theta_4}},\label{s-sol}
\end{align}
with $\theta_j=k_jx+\omega_j t+\delta_j$, $j=1,2,3,4$ satisfying the dispersion relations (\ref{dispersion}).

\section{Solutions of two-place shifted nonlocal equations}

Here we present examples of soliton solutions of two-place shifted nonlocal equations.

\noindent \textbf{Example 1.} Consider the equation (\ref{twopex1}) which is derived from the AKNS system (\ref{4NLS-a})-(\ref{4NLS-d}) by the reduction formulas
$p(x,t)=\bar{q}(x,-t+t_0)$, $r(x,t)=\bar{q}(x,t)$, $s(x,t)=q(x,-t+t_0)$, and $c=-\bar{c}$. We use these reduction formulas with the solution (\ref{q-sol})-(\ref{s-sol}), and obtain the following constraints:
\begin{align}
&k_2=k_3=\bar{k}_1, \quad k_4=k_1,\quad \omega_2=-\bar{\omega}_1,\quad \omega_3=\bar{\omega}_1,\quad \omega_4=-\omega_1,\\
&e^{\delta_2}=e^{\bar{\delta}_1+\bar{\omega}_1t_0},\quad e^{\delta_3}=e^{\bar{\delta}_1},\quad e^{\delta_4}=e^{\delta_1+\omega_1t_0}.
\end{align}
Hence one-soliton solution of the two-place shifted time reversal equation (\ref{twopex1}) is
{\small\begin{equation}\label{ex1sol}
q(x,t)=\frac{e^{k_1x+\omega_1t+\delta_1}+\frac{\sigma(k_1-\bar{k}_1)}{2k_1(k_1+\bar{k}_1)^2}e^{(2k_1+\bar{k}_1)x-\bar{\omega}_1t+2\delta_1+\bar{\delta}_1
+(\omega_1+\bar{\omega}_1)t_0}}{1+\frac{\sigma e^{(k_1+\bar{k}_1)x+\delta_1+\bar{\delta}_1}}
{(k_1+\bar{k}_1)^2}[e^{(\omega_1+\bar{\omega}_1)t}+e^{(\omega_1+\bar{\omega}_1)(-t+t_0) }]-\frac{(k_1-\bar{k}_1)^2}{4k_1\bar{k}_1(k_1+\bar{k}_1)^4}
e^{2(k_1+\bar{k}_1)x+2(\delta_1+\bar{\delta}_1)+(\omega_1+\bar{\omega}_1)t_0  }     }.
\end{equation}}
Here $k_1+\bar{k}_1\neq 0$. The above solution is nonsingular if the denominator, say $W(x,t)$, does not vanish. Let $k_1=a_1+ib_1$, $c=i\xi$, $\delta_1=\alpha_1+i\beta_1$. The denominator becomes
\begin{equation}
W(x,t)=1+\sigma \frac{1}{2a_1^2}Y\cosh(\theta)+\frac{b_1^2}{16a_1^4(a_1^2+b_1^2)}Y^2,
\end{equation}
where
\begin{equation}
Y=e^{2a_1x+2\alpha_1+\frac{st_0}{2}},\,\, \theta=s\Big(t-\frac{t_0}{2}\Big),\,\, s=\omega_1+\bar{\omega}_1=-\frac{4a_1b_1}{\xi},\,\, a_1\neq 0.
\end{equation}
Note that $Y>0$ and $\cosh(\theta)\geq 1$. Hence $W(x,t)$ is not zero, i.e., the solution (\ref{ex1sol}) is nonsingular if $\sigma=1$.

Let us choose particular values of parameters; $\sigma=1$, $c=i$, $k_1=1+i$, $\delta_1=0$, $t_0=2$. Then the real-valued solution $|q|^2$ of the equation (\ref{twopex1})
becomes
\begin{equation}
|q|^2=\frac{32e^{2x+4t}(8e^{2x+4t-8}+e^{4x+8t-16}+32)}{[32e^{4t}+8e^{2x}+8e^{2x+8t-8}+e^{4x+4t-8}]^2}.
\end{equation}
The 3D and contour plot graphs of the above solution $|q|^2$ are given in the following Figure 1.
\begin{center}
\begin{figure}[h!]
\centering
\subfloat[]{\includegraphics[width=0.30\textwidth]{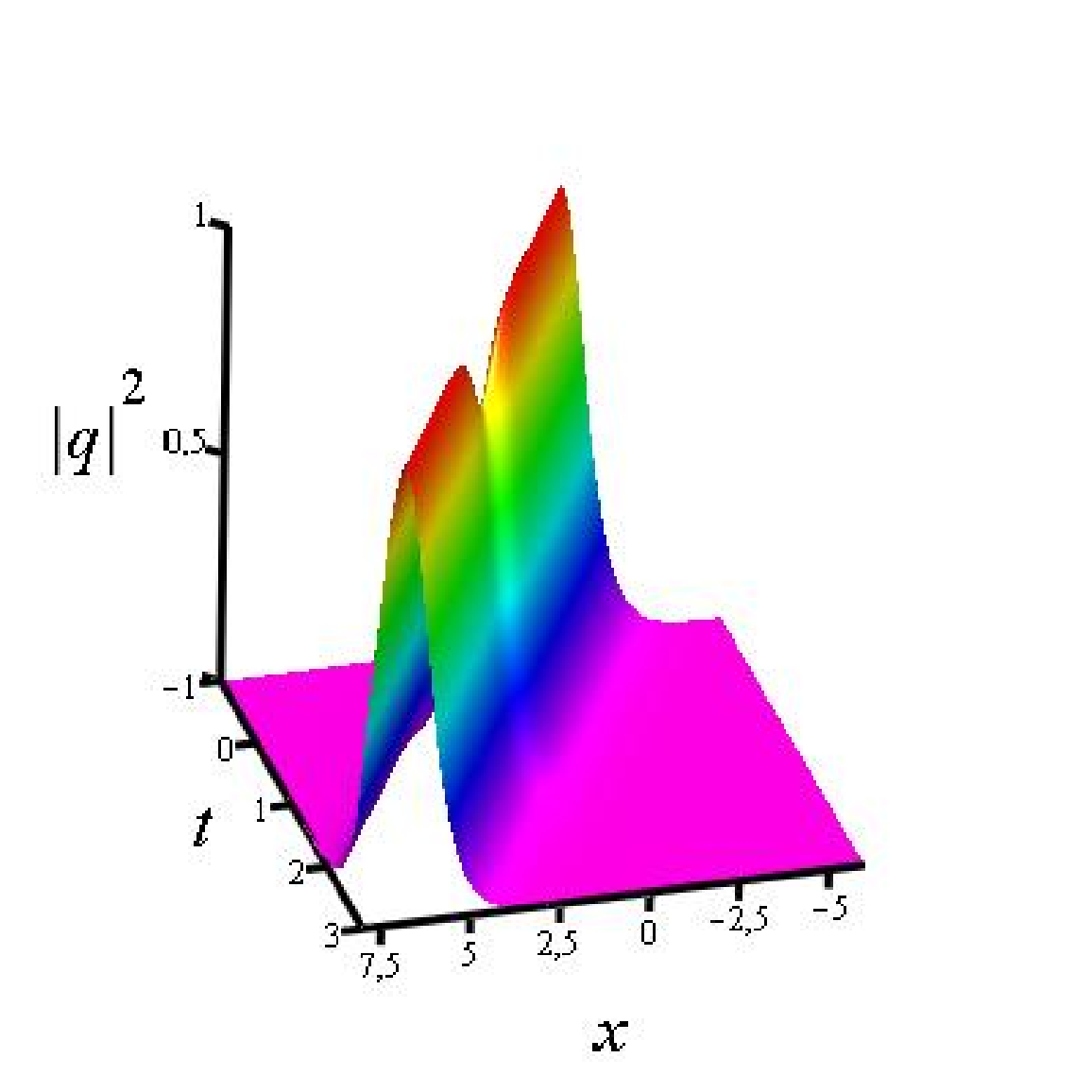}}\hspace{2cm}
\subfloat[]{\includegraphics[width=0.30\textwidth]{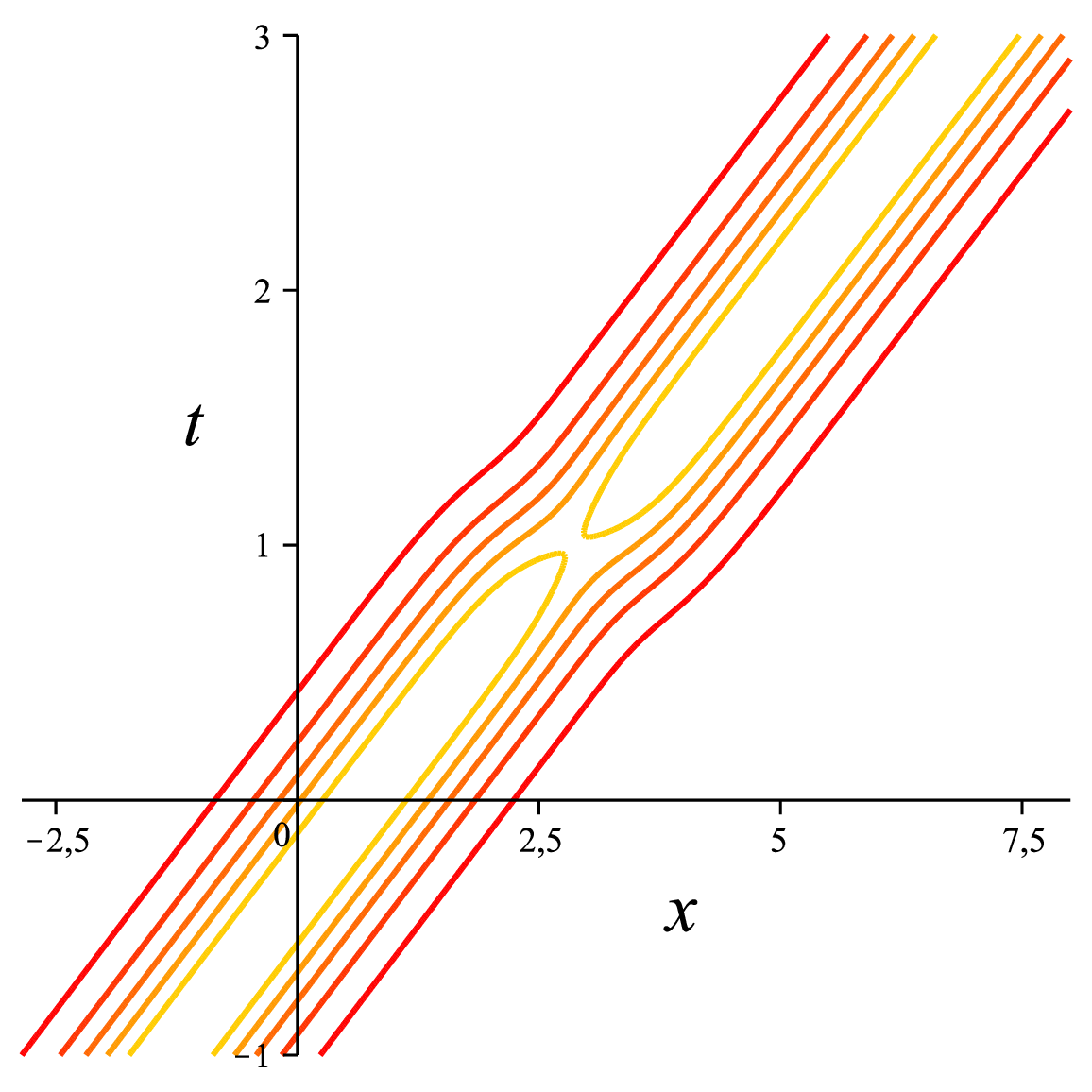}}
\caption{One-soliton solution $|q|^2$ of the equation (\ref{twopex1}) (a) 3D graph, (b) contour plot.}
\end{figure}
\end{center}
\squeezeup

\noindent \textbf{Example 2.} Take the equation (\ref{twopex2}) which is obtained from (\ref{4NLS-a})-(\ref{4NLS-d}) by the reduction formulas $p(x,t)=q(-x+x_0,t)$, $r(x,t)=\bar{q}(x,t)$, $s(x,t)=\bar{q}(-x+x_0,t)$, and $c=-\bar{c}$. Using these reductions with the solution (\ref{q-sol})-(\ref{s-sol}) we get
\begin{align}
&k_2=-k_1,\quad k_3=\bar{k}_1,\quad k_4=-\bar{k}_1,\quad \omega_2=\omega_1,\quad \omega_3=\bar{\omega}_1,\quad \omega_4=\bar{\omega}_1,\\
&e^{\delta_2}=e^{\delta_1+k_1x_0},\quad e^{\delta_3}=e^{\bar{\delta}_1},\quad e^{\delta_4}=e^{\bar{\delta}_1+\bar{k}_1x_0}.
\end{align}
Therefore one-soliton solution of the equation (\ref{twopex2})  is
{\small \begin{equation}\label{ex2sol}
q(x,t)=\frac{e^{k_1x+\omega_1t+\delta_1}+\frac{2\sigma k_1}{(k_1-\bar{k}_1)(k_1+\bar{k}_1)^2}e^{-\bar{k}_1x+(2\omega_1+\bar{\omega}_1)t+2\delta_1+\bar{\delta}_1+(k_1+\bar{k}_1)x_0}            }{1+\frac{\sigma e^{(\omega_1+\bar{\omega}_1)t+\delta_1+\bar{\delta}_1}}{(k_1+\bar{k}_1)^2}[e^{(k_1+\bar{k}_1)x}+e^{-(k_1+\bar{k}_1)x+(k_1+\bar{k}_1)x_0}]
-\frac{4k_1\bar{k}_1e^{2(\omega_1+\bar{\omega}_1)t+2(\delta_1+\bar{\delta}_1)+(k_1+\bar{k}_1)x_0}}{(k_1-\bar{k}_1)^2(k_1+\bar{k}_1)^4}        }.
\end{equation}}
Here $k_1+\bar{k}_1\neq 0$ and $k_1-\bar{k}_1\neq 0$. The solution (\ref{ex2sol}) is singular when its denominator, say $W(x,t)$, vanishes.  Let $k_1=a+ib$, $\delta_1=\alpha+i\beta$, and $c=i\xi$. The denominator
$W(x,t)$ becomes
\begin{equation}
W(x,t)=1+\frac{\sigma}{2a^2}e^{\phi+ax_0}\cosh\Big(2a\Big(x-\frac{x_0}{2}\Big)\Big)+\frac{a^2+b^2}{16a^4b^2}e^{2\phi+2ax_0},
\end{equation}
where $\phi=-\frac{4ab}{\xi}t+2\alpha$. For $\sigma=1$, $W(x,t)>0$ and the solution (\ref{ex2sol}) is nonsingular.

Choosing the parameters of the solution (\ref{ex2sol}) as $\sigma=x_0=\delta_1=1$, $c=i$, $k_1=1+i$. Then we obtain the real-valued solution $|q|^2$ of the equation (\ref{twopex2}) as
{\small\begin{align}
&|q|^2=(8\left(4e^{8t+4x+6}+8e^{12t+6x+2}+e^{4t+2x+10}\right))\Big/\Big(64e^{16t+4x}+32e^{12t+6x+2}+32e^{12t+2x+4}\nonumber\\
&+24e^{8t+4x+6}
+4e^{8t+8x+4}+4e^{4t+6x+8}+4e^{8t+8}+4e^{4t+2x+10}+e^{4x+12}\Big).
\end{align}}
The 3D and contour plot graphs of the above solution $|q|^2$ are given in the following Figure 2.
\begin{center}
\begin{figure}[h!]
\centering
\subfloat[]{\includegraphics[width=0.30\textwidth]{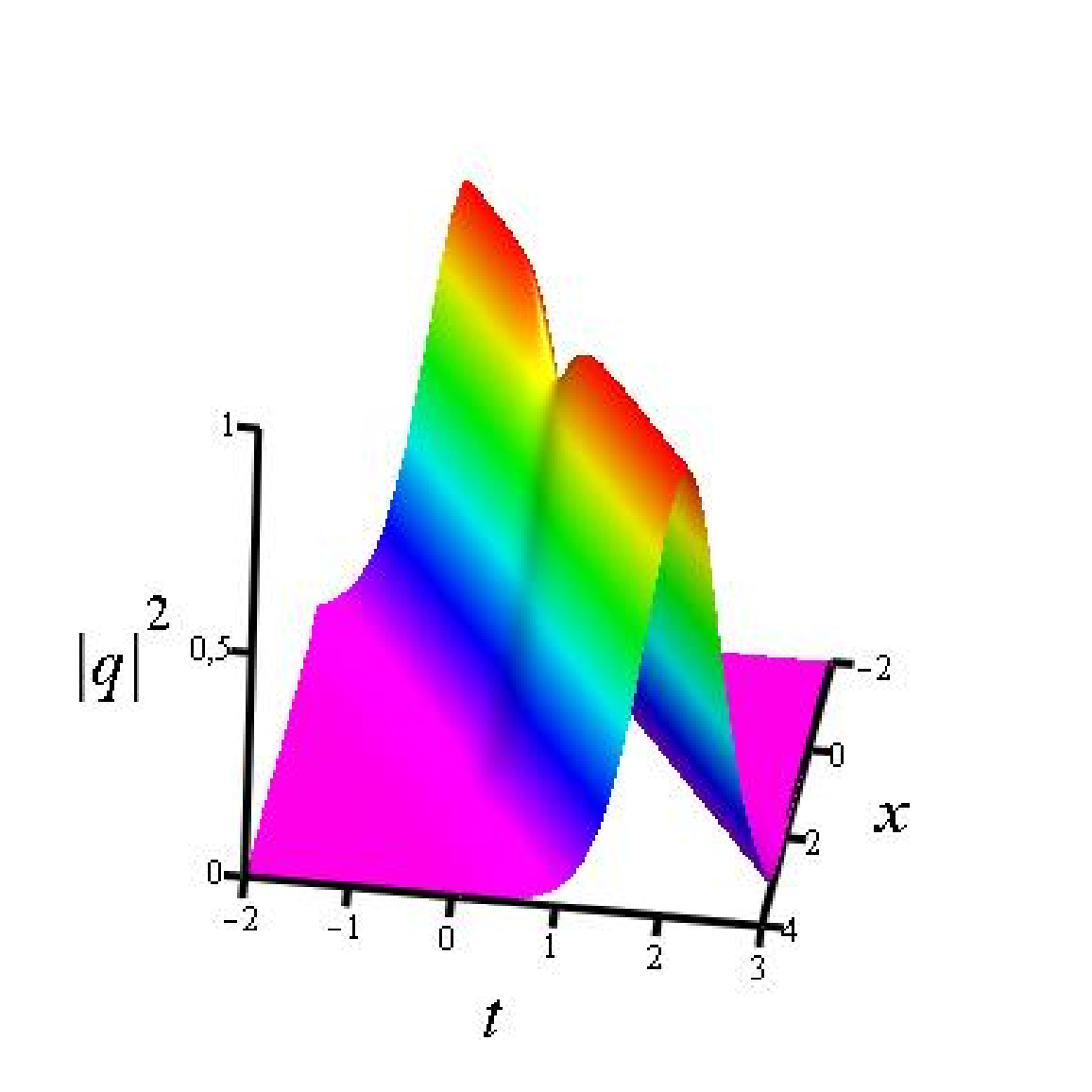}}\hspace{2cm}
\subfloat[]{\includegraphics[width=0.30\textwidth]{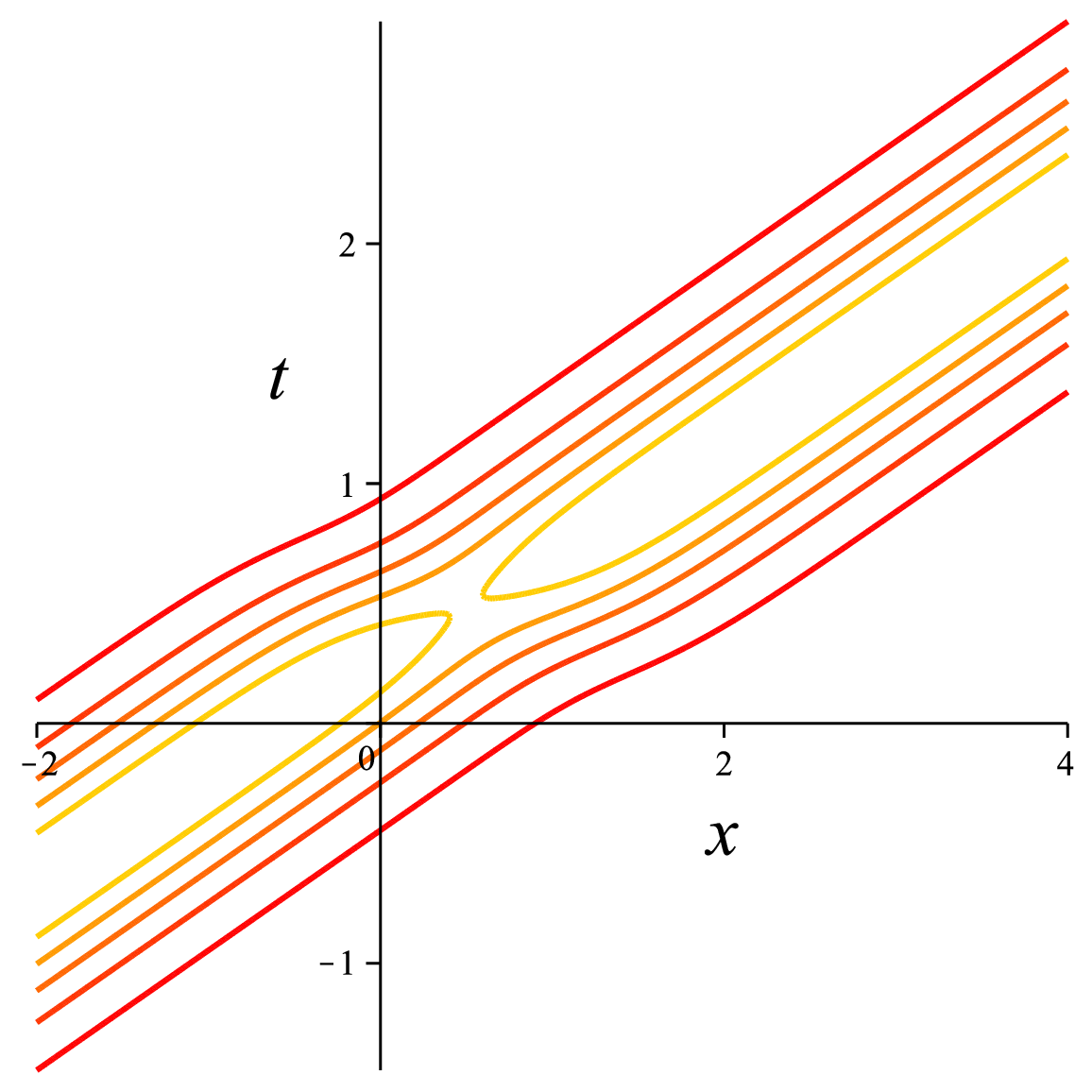}}
\caption{One-soliton solution $|q|^2$ of the equation (\ref{twopex2}) (a) 3D graph, (b) contour plot.}
\end{figure}
\end{center}
\squeezeup

\noindent \textbf{Example 3.} Consider the equation (\ref{twopex3}) which is obtained from (\ref{4NLS-a})-(\ref{4NLS-d}) by the reduction formulas $p(x,t)=q(-x+x_0,t)$, $r(x,t)=\bar{q}(-x+x_0,t)$, $s(x,t)=\bar{q}(x,t)$, and $c=-\bar{c}$. Using these reductions with the solution (\ref{q-sol})-(\ref{s-sol}) we have
\begin{align}
&k_2=-k_1,\quad k_3=-\bar{k}_1,\quad k_4=\bar{k}_1,\quad \omega_2=\omega_1,\quad \omega_3=\bar{\omega}_1,\quad \omega_4=\bar{\omega}_1,\\
&e^{\delta_2}=e^{\delta_1+k_1x_0},\quad e^{\delta_3}=e^{\bar{\delta}_1+\bar{k}_1x_0},\quad e^{\delta_4}=e^{\bar{\delta}_1}.
\end{align}
Hence one-soliton solution of the equation (\ref{twopex3})  is
{\small\begin{equation}\label{ex3sol}
q(x,t)=\frac{e^{k_1x+\omega_1t+\delta_1}+\frac{2\sigma k_1}{(k_1+\bar{k}_1)(k_1-\bar{k}_1)^2}e^{\bar{k}_1x+(2\omega_1+\bar{\omega}_1)t+2\delta_1+\bar{\delta}_1+k_1x_0}            }{1+\frac{\sigma e^{(\omega_1+\bar{\omega}_1)t+\delta_1+\bar{\delta}_1+\bar{k}_1x_0}}{(k_1-\bar{k}_1)^2}[e^{(k_1-\bar{k}_1)x}+e^{-(k_1-\bar{k}_1)x+(k_1-\bar{k}_1)x_0}]
+\frac{4k_1\bar{k}_1e^{2(\omega_1+\bar{\omega}_1)t+2(\delta_1+\bar{\delta}_1)+(k_1+\bar{k}_1)x_0}}{(k_1+\bar{k}_1)^2(k_1-\bar{k}_1)^4}        }.
\end{equation}}
Here $k_1+\bar{k}_1\neq 0$ and $k_1-\bar{k}_1\neq 0$. For the singularity analysis of the solution (\ref{ex3sol}) we check the denominator, say $W(x,t)$, of this solution. Let $k_1 = a + ib$ for $a \neq 0$, $b \neq 0$, $\delta_1 = \alpha + i\beta$, and $c = i\xi$. Then the denominator $W(x,t)$ becomes
\begin{equation}
W(x,t) = 1 - \frac{\sigma}{2} Y(t)\cos(2bx - bx_0) + \frac{a^2 + b^2}{16a^2} Y^2(t),
\quad
Y(t) = \frac{e^{-\frac{4ab}{\xi}t + 2\alpha + ax_0}}{b^2}.
\end{equation}
Letting $A=\frac{a^2 + b^2}{16a^2} > 0$ and $\gamma=\frac{\sigma}{2} \cos(2bx - bx_0)  \in [-\frac{1}{2},\frac{1}{2}]$ we have $W(x,t) = A Y^2 - \gamma Y + 1$. Now the question is whether $W(x,t) = 0$ for some $(x,t)$. We check the discriminant $\Delta = \gamma^2 - 4A = \gamma^2 - \frac{a^2 + b^2}{4a^2}$. Since $\gamma^2 \leq \frac{1}{4}$ we have $\Delta\leq \frac{1}{4}-\frac{a^2 + b^2}{4a^2}=-\frac{b^2}{4a^2}<0$ with $A > 0$. Therefore $W(x,t) > 0$ for all $x,t$, so the solution (\ref{ex3sol}) of the equation (\ref{twopex3}) is nonsingular for all $x,t$ if $a \neq 0$, $b \neq 0$, $\xi \neq 0$.

Take the parameters of the solution (\ref{ex3sol}) as $\sigma=1$, $c=i$, $k_1=1+i$, $\delta_1=1-i$, $x_0=1$. So we obtain the real-valued solution $|q|^2$ of the equation (\ref{twopex3}) as
\begin{equation}
|q|^2=\frac{8e^{2x+4t}[4e^{4t+5}\cos(2x-1)+4e^{4t+5}\sin(2x-1)-8e^{8t+2}-e^8] }{8\cos(2x-1)[e^{4t+9}+8e^{12t+3}]-8e^{8t+6}\cos(4x-2)-24e^{8t+6}-64e^{16t}-e^{12}}.
\end{equation}
The 3D and contour plot graphs of the above solution $|q|^2$ are given in the following Figure 3.
\begin{center}
\begin{figure}[h!]
\centering
\subfloat[]{\includegraphics[width=0.30\textwidth]{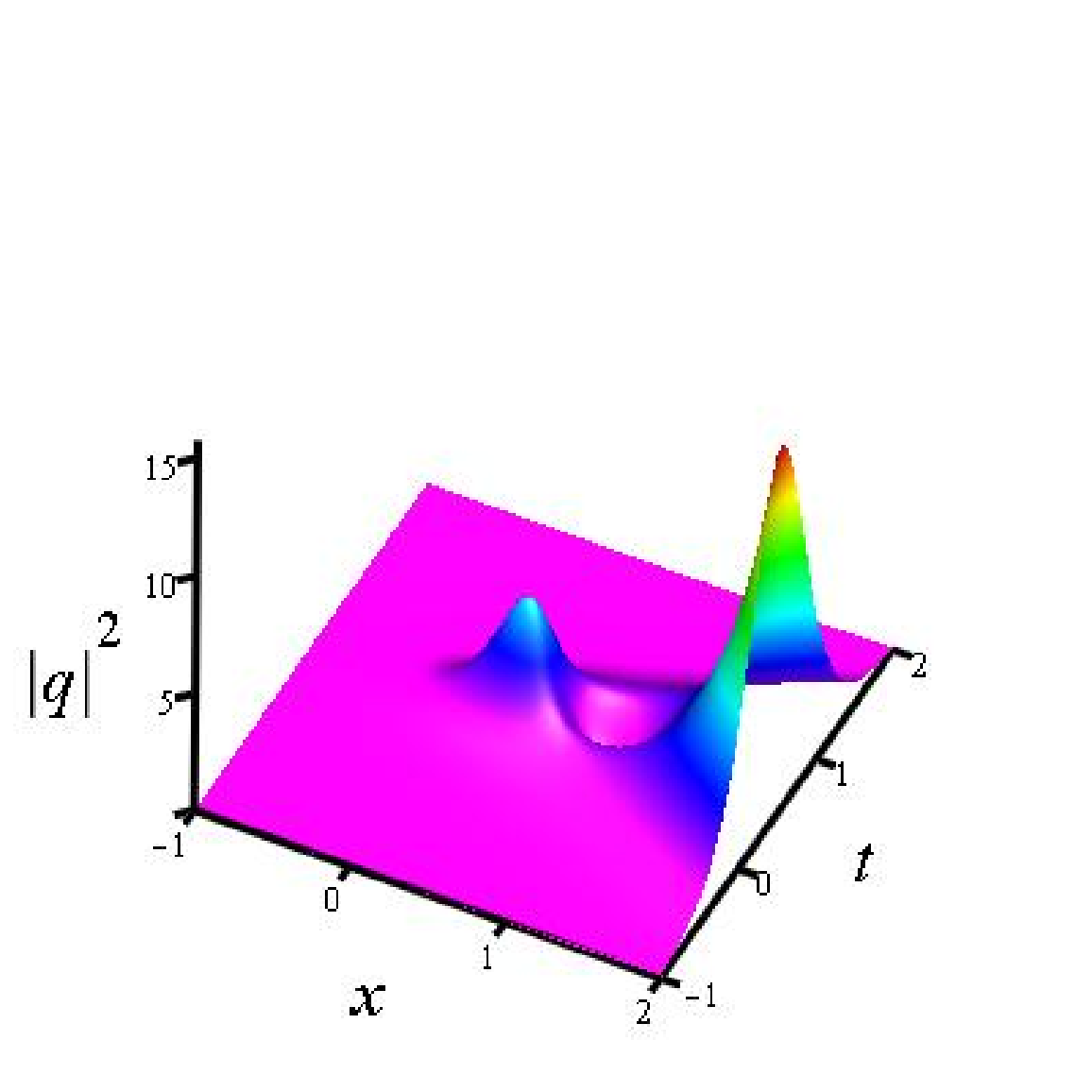}}\hspace{2cm}
\subfloat[]{\includegraphics[width=0.30\textwidth]{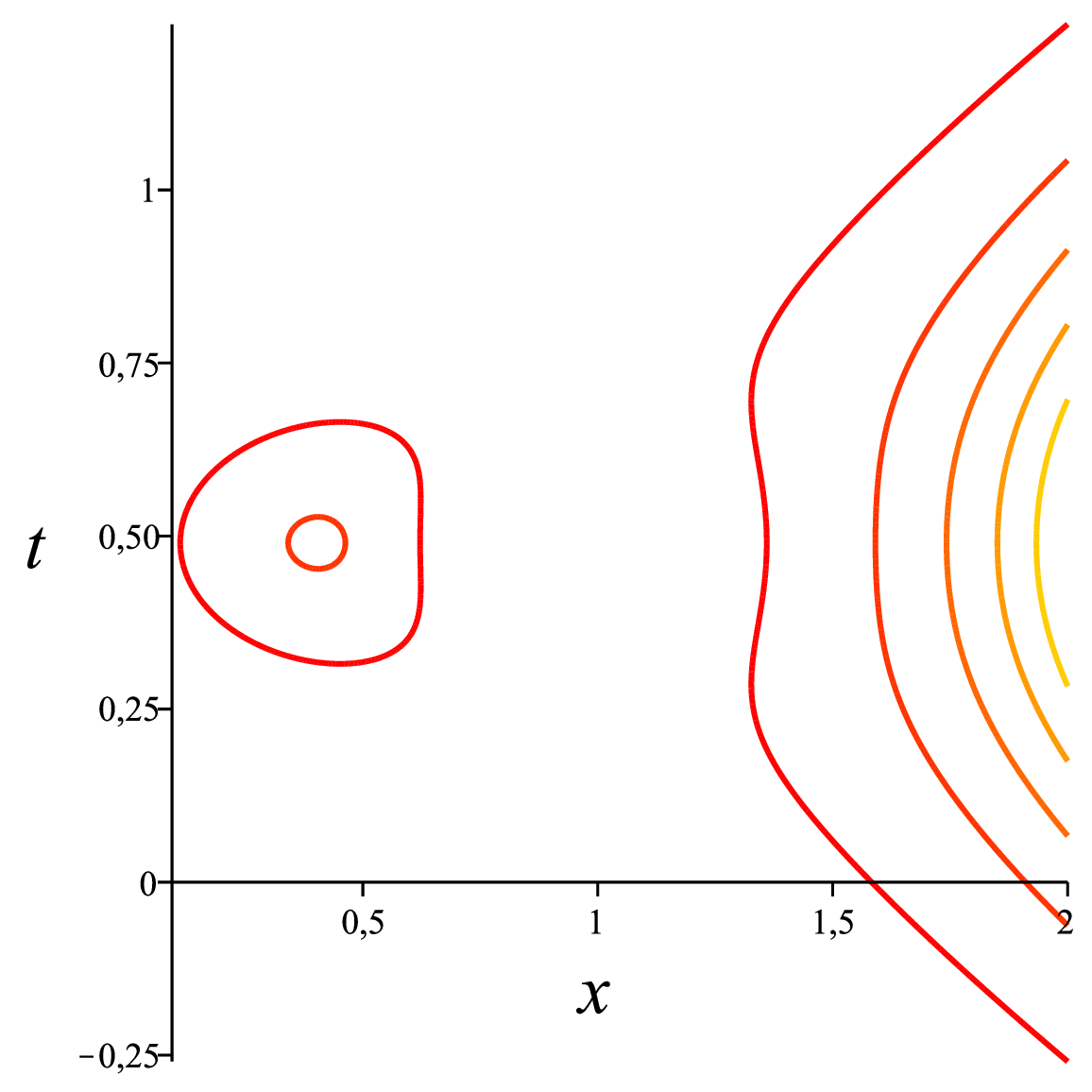}}
\caption{One-soliton solution $|q|^2$ of the equation (\ref{twopex3}) (a) 3D graph, (b) contour plot.}
\end{figure}
\end{center}
\squeezeup

\noindent \textbf{Example 4.} Take the equation (\ref{twopex4}) which is derived from (\ref{4NLS-a})-(\ref{4NLS-d}) by the reduction formulas $p(x,t)=q(x,t)$, $r(x,t)=q(x,-t+t_0)$, $s(x,t)=q(x,-t+t_0)$. By our method, we obtain
\begin{align}
k_1=k_2=k_3=k_4,\quad \omega_2=\omega_1,\quad \omega_3=\omega_4=-\omega_1,\quad e^{\delta_2}=e^{\delta_1},\quad e^{\delta_3}=e^{\delta_4}=e^{\delta_1+\omega_1t_0}.
\end{align}
So we obtain the one-soliton solution of the equation (\ref{twopex4})  as
\begin{equation}\label{ex4sol}
q(x,t)=\frac{e^{k_1x+\omega_1t+\delta_1}}{1+\frac{\sigma}{2k_1^2}e^{2k_1x+2\delta_1+\omega_1t_0}}.
\end{equation}
Here $k_1\neq 0$. If we let $k_1, c, \delta_1 \in \mathbb{R}$, then the solution (\ref{ex4sol}) is real and it is nonsingular for $\sigma=1$. For a particular example, take $\sigma=c=1$, $k_1=\frac{1}{2}$, $\delta_1=-1$, $t_0=2$. Hence one-soliton solution of the equation (\ref{twopex4}) is
\begin{equation}\displaystyle
q=\frac{e^{\frac{1}{2}x-\frac{1}{4}t-1}}{1+2e^{x-\frac{5}{2}}}.
\end{equation}
The 3D and contour plot graphs of the above solution are given in the following Figure 4.
\begin{center}
\begin{figure}[h!]
\centering
\subfloat[]{\includegraphics[width=0.30\textwidth]{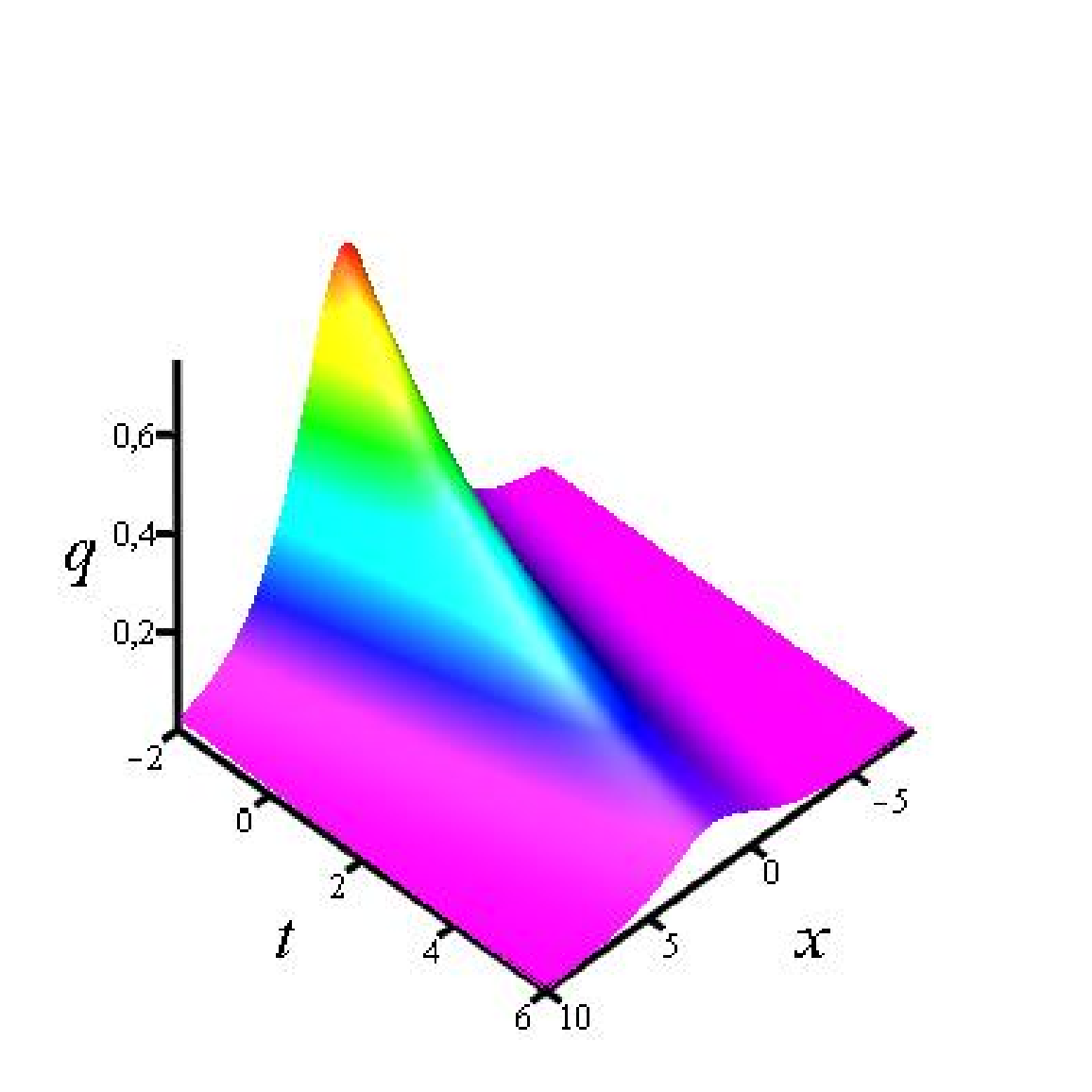}}\hspace{2cm}
\subfloat[]{\includegraphics[width=0.30\textwidth]{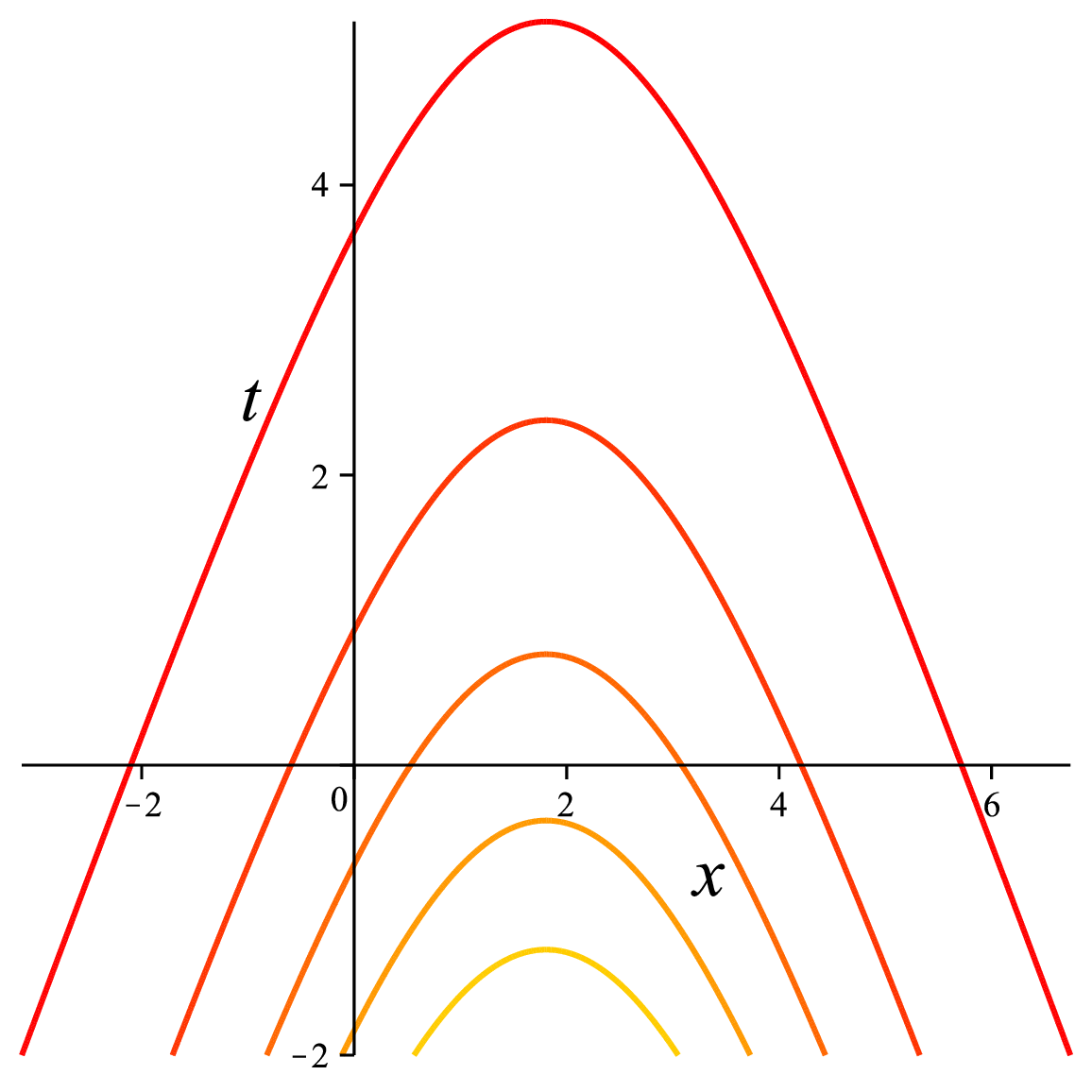}}
\caption{Asymptotically decaying solution $q$ of the equation (\ref{twopex4}) (a) 3D graph, (b) contour plot.}
\end{figure}
\end{center}
\squeezeup

\noindent \textbf{Example 5.} Consider the equation (\ref{twopex6}) which is derived from (\ref{4NLS-a})-(\ref{4NLS-d}) by the reduction formulas  $p(x,t)=q(x,t)$, $r(x,t)=\bar{q}(x,-t+t_0)$, $s(x,t)=\bar{q}(x,-t+t_0)$, and $c=\bar{c}$. From our solution method we have
\begin{align}
k_2=k_1,\quad k_3=k_4=\bar{k}_1,\quad \omega_2=\omega_1,\quad \omega_3=\omega_4=-\bar{\omega}_1,\quad e^{\delta_2}=e^{\delta_1},\quad e^{\delta_3}=e^{\delta_4}=e^{\bar{\delta}_1+\bar{\omega}_1t_0}.
\end{align}
Therefore we have the one-soliton solution of the equation (\ref{twopex6})  as
\begin{equation}\label{ex5sol}
q(x,t)=\frac{e^{k_1x+\omega_1t+\delta_1}}{1+\frac{2\sigma}{(k_1+\bar{k}_1)^2}e^{(k_1+\bar{k}_1)x+(\omega_1-\bar{\omega}_1)t+\delta_1+\bar{\delta}_1+\bar{\omega}_1t_0}}.
\end{equation}
Here $k_1+\bar{k}_1\neq 0$. Let us analyze the singularity of the above solution. If the denominator of (\ref{ex5sol}), say $W(x,t)$, is zero for some $(x,t)$, then the solution is singular.
Letting $k_1=a+ib$, $a\neq 0$, and $\delta_1=\alpha+i\beta$. Then
\begin{equation}\displaystyle
W(x,t)=1+\frac{\sigma}{2a^2}e^{2ax+2\alpha+\Big( \frac{(b^2-a^2)}{c} \Big)t_0}e^{-\frac{2iab}{c}( 2t-t_0)}=0
\end{equation}
for
\begin{equation}
\frac{1}{2a^2}e^{2ax+2\alpha+\Big( \frac{(b^2-a^2)}{c} \Big)t_0}=1,\quad \sigma e^{-\frac{2iab}{c}( 2t-t_0)}=-1.
\end{equation}
Assume that $b\neq 0$. Hence the solution (\ref{ex5sol}) is singular at $t=\frac{t_0}{2}-\frac{c(2n+1)\pi}{4ab}$ if $\sigma=1$, and
 at $t=\frac{t_0}{2}-\frac{cn\pi}{2ab}$ if $\sigma=-1$ for $x=\frac{1}{2a}(\ln(2a^2)-2\alpha+\frac{(a^2-b^2)}{c}t_0)$.

 If $b=0$, we have the denominator just depending on $x$ and
 \begin{equation}
 W(x)=1+\frac{\sigma}{2a^2}e^{2ax+2\alpha-\frac{a^2}{c}t_0}\in \mathbb{R}.
 \end{equation}
Therefore, in this case for $\sigma=1$, the solution (\ref{ex5sol}) of the equation (\ref{twopex6}) is nonsingular. Choosing the parameters of the solution (\ref{ex5sol})
as $c=\sigma=k_1=\delta_1=t_0=1$ yields
\begin{equation}\displaystyle
q=\frac{e^{x-t+1}}{1+\frac{1}{2}e^{2x+1}}.
\end{equation}
The 3D and contour plot graphs of the above solution are given in the following Figure 5.
\begin{center}
\begin{figure}[h!]
\centering
\subfloat[]{\includegraphics[width=0.30\textwidth]{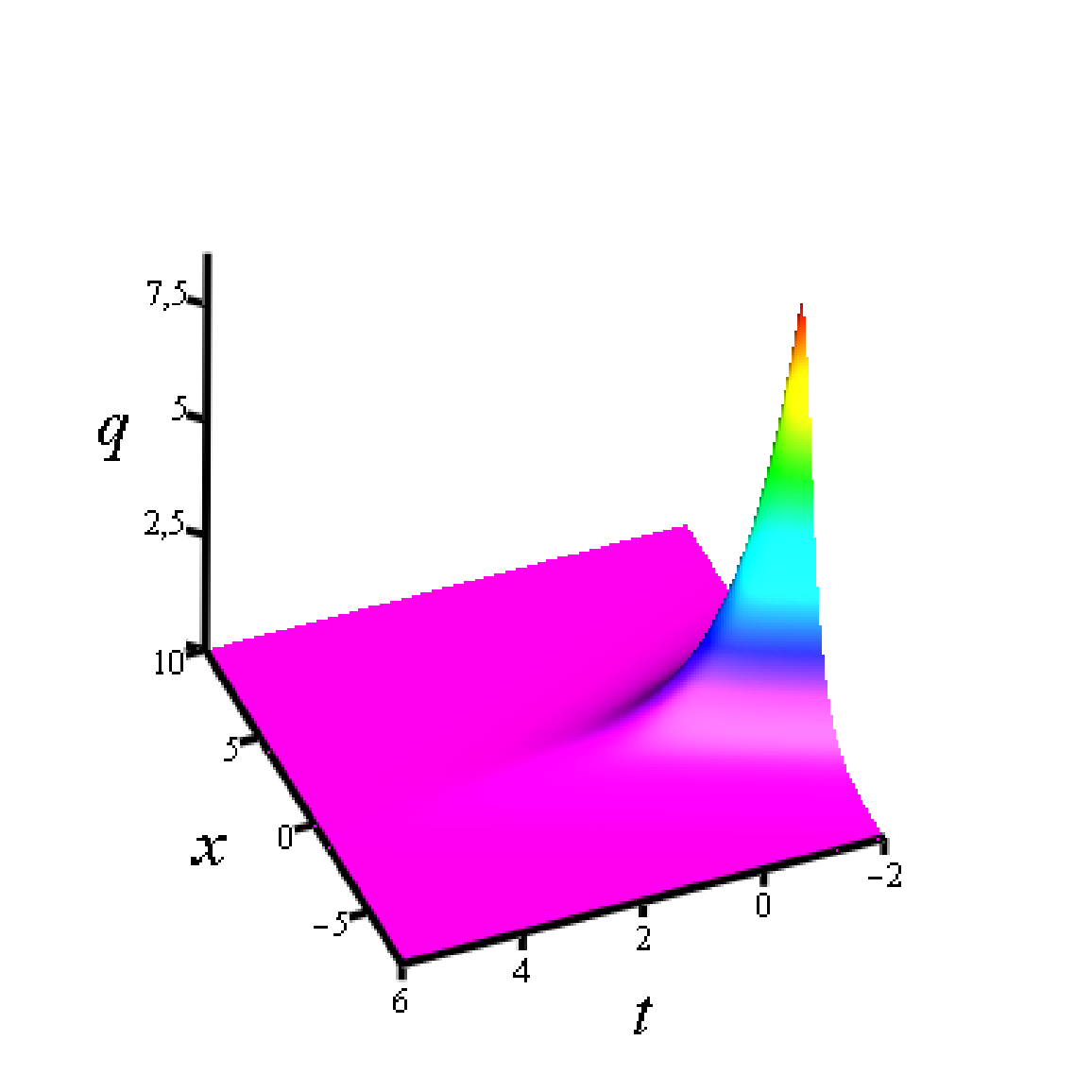}}\hspace{2cm}
\subfloat[]{\includegraphics[width=0.30\textwidth]{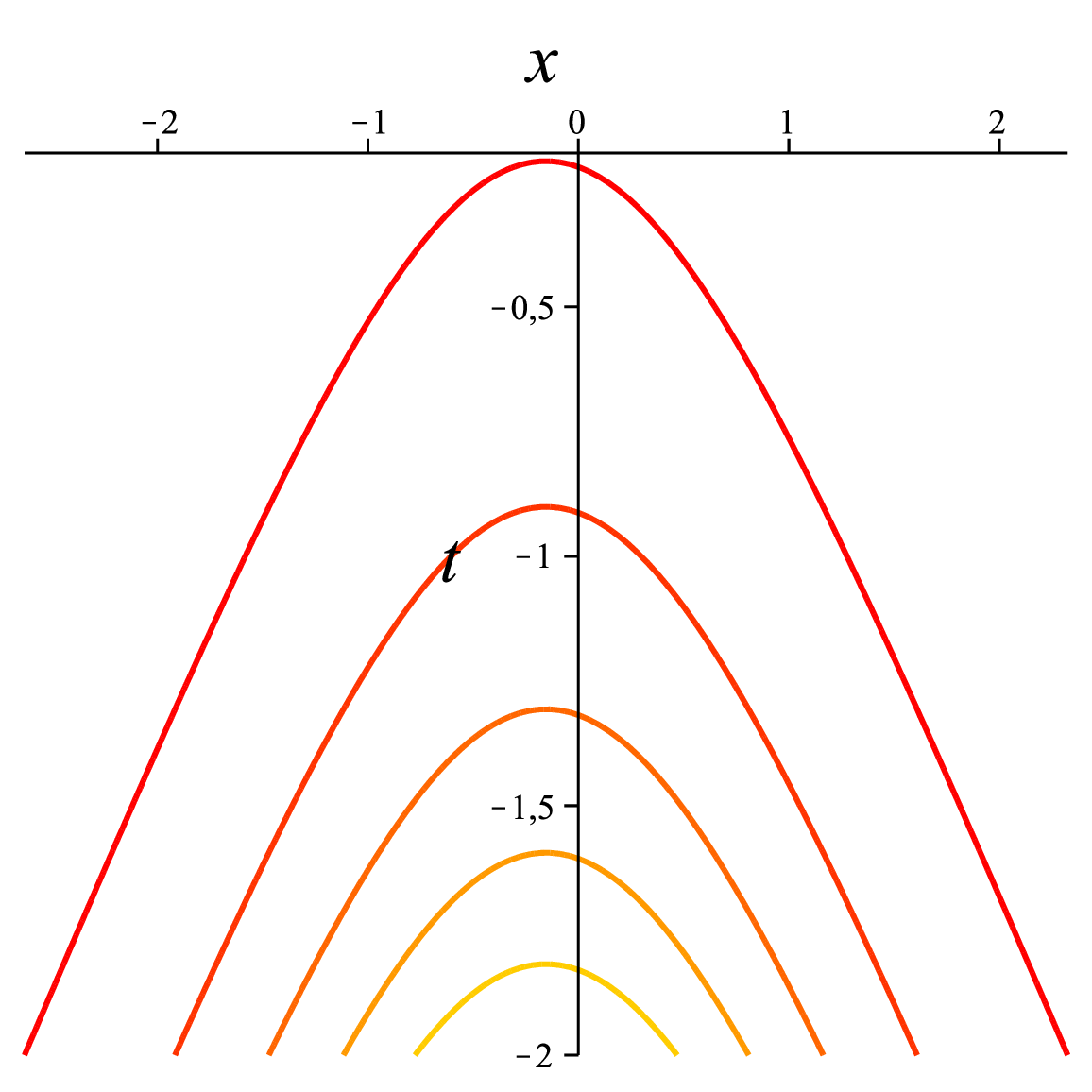}}
\caption{Asymptotically decaying solution $q$ of the equation (\ref{twopex6}) (a) 3D graph, (b) contour plot.}
\end{figure}
\end{center}
\squeezeup

\noindent \textbf{Example 6.} Take the equation (\ref{twopex7}) which is derived from (\ref{4NLS-a})-(\ref{4NLS-d}) by the reduction formulas  $p(x,t)=q(x,t)$, $r(x,t)=\bar{q}(-x+x_0,t)$, $s(x,t)=\bar{q}(-x+x_0,t)$, and $c=-\bar{c}$. We use these reductions with the solution (\ref{q-sol})-(\ref{s-sol}) and  get
\begin{equation}
k_2=k_1,\,\,\, k_3=k_4=-\bar{k}_1,\,\,\, \omega_2=\omega_1,\,\,\, \omega_3=\omega_4=\bar{\omega}_1,\,\,\, e^{\delta_2}=e^{\delta_1},\,\,\, e^{\delta_3}=e^{\delta_4}=e^{\bar{\delta}_1+\bar{k}_1x_0}.
\end{equation}
Hence we obtain the one-soliton solution of the equation (\ref{twopex7})  as
\begin{equation}\label{ex6sol}
q(x,t)=\frac{e^{k_1x+\omega_1t+\delta_1}}{1+\frac{2\sigma}{(k_1-\bar{k}_1)^2}e^{(k_1-\bar{k}_1)x+(\omega_1+\bar{\omega}_1)t+\delta_1+\bar{\delta}_1+\bar{k}_1x_0}}.
\end{equation}
Here $k_1-\bar{k}_1\neq 0$. The solution (\ref{ex6sol}) is singular when its denominator, say $W(x,t)$, vanishes at some $(x,t)$. Let $c=i\xi$, $k_1=a+ib$ $(b\neq 0)$, $\delta_1=\alpha+i\beta$.
We have
\begin{equation}
W(x,t)=1-\frac{\sigma}{2b^2}e^{-\frac{4ab}{\xi}t+2\alpha+ax_0+i(2bx-bx_0)}.
\end{equation}
It becomes zero if
\begin{equation}
\frac{1}{2b^2}e^{2\alpha+ax_0-\frac{4ab}{\xi}t}=1,\quad \sigma e^{i(2bx-bx_0)}=1.
\end{equation}
Assume that $ab\neq 0$. Hence the solution (\ref{ex6sol}) is singular at $x=\frac{1}{2b}(2n\pi+bx_0)$ if $\sigma=1$, and at $x=\frac{1}{2b}((2n+1)\pi+bx_0)$ if $\sigma=-1$
with $t=\frac{\xi}{4ab}(2\alpha+ax_0-\ln(2b^2))$.

If $a=0$ then the denominator becomes a function of $x$ only,
\begin{equation}
W(x)=1-\frac{\sigma}{2b^2}e^{i(2bx-bx_0)+2\alpha}.
\end{equation}
If $e^{2\alpha}\neq 2b^2$ then $W(x)\neq 0$, so the solution (\ref{ex6sol}) is nonsingular if $a=0$, $b\neq 0$, and  $e^{2\alpha}\neq 2b^2$.
Choose the parameters of the solution (\ref{ex6sol}) as $k_1=\frac{3i}{2}, c=i, x_0=2, \delta_1=1-i, \sigma=-1$ giving
\begin{equation}\displaystyle
|q|^2=\frac{81e^2}{81+36e^2\cos(3x-3)+4e^4}.
\end{equation}
The 3D and contour plot graphs of the above solution are given in the following Figure 6.
\begin{center}
\begin{figure}[h!]
\centering
\subfloat[]{\includegraphics[width=0.30\textwidth]{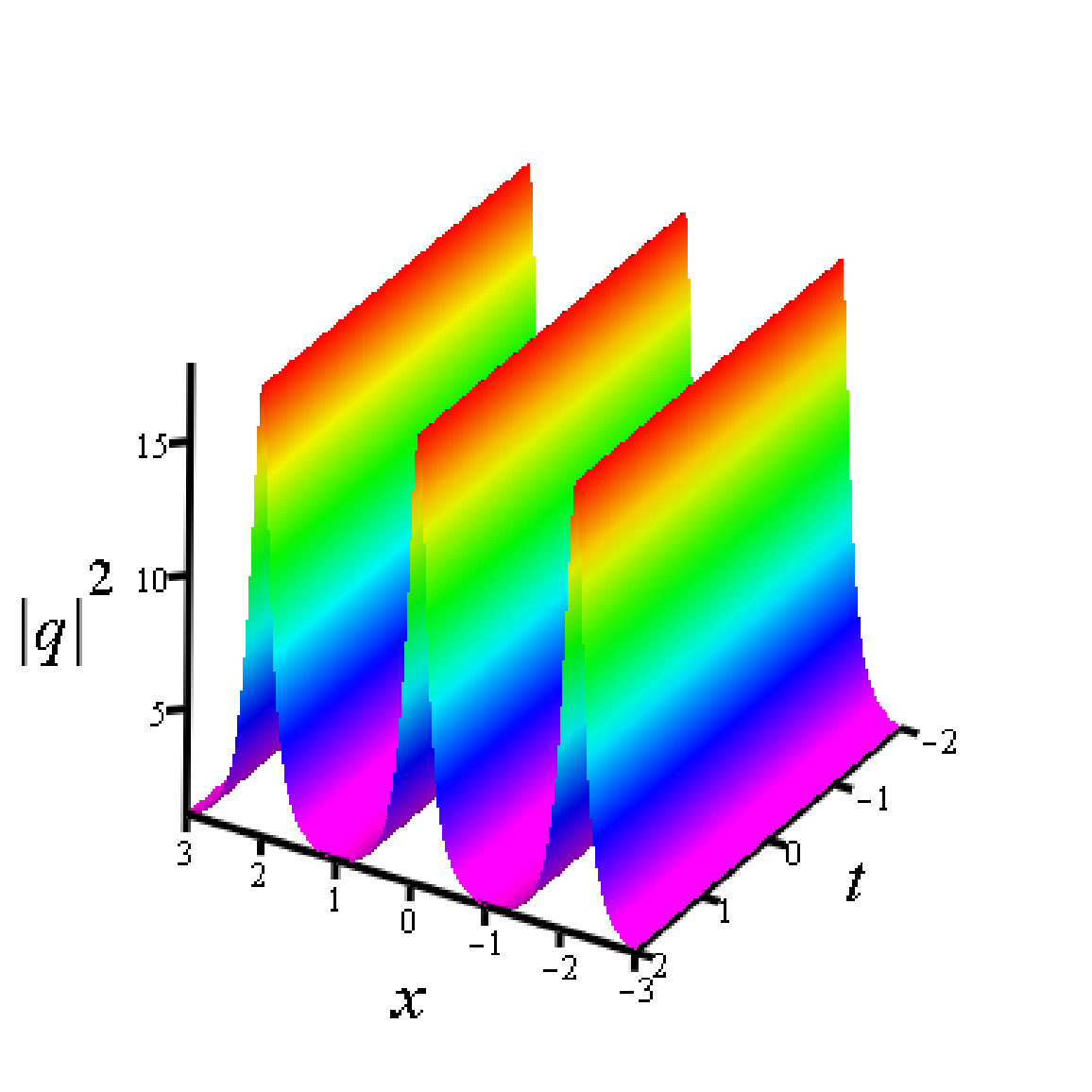}}\hspace{2cm}
\subfloat[]{\includegraphics[width=0.30\textwidth]{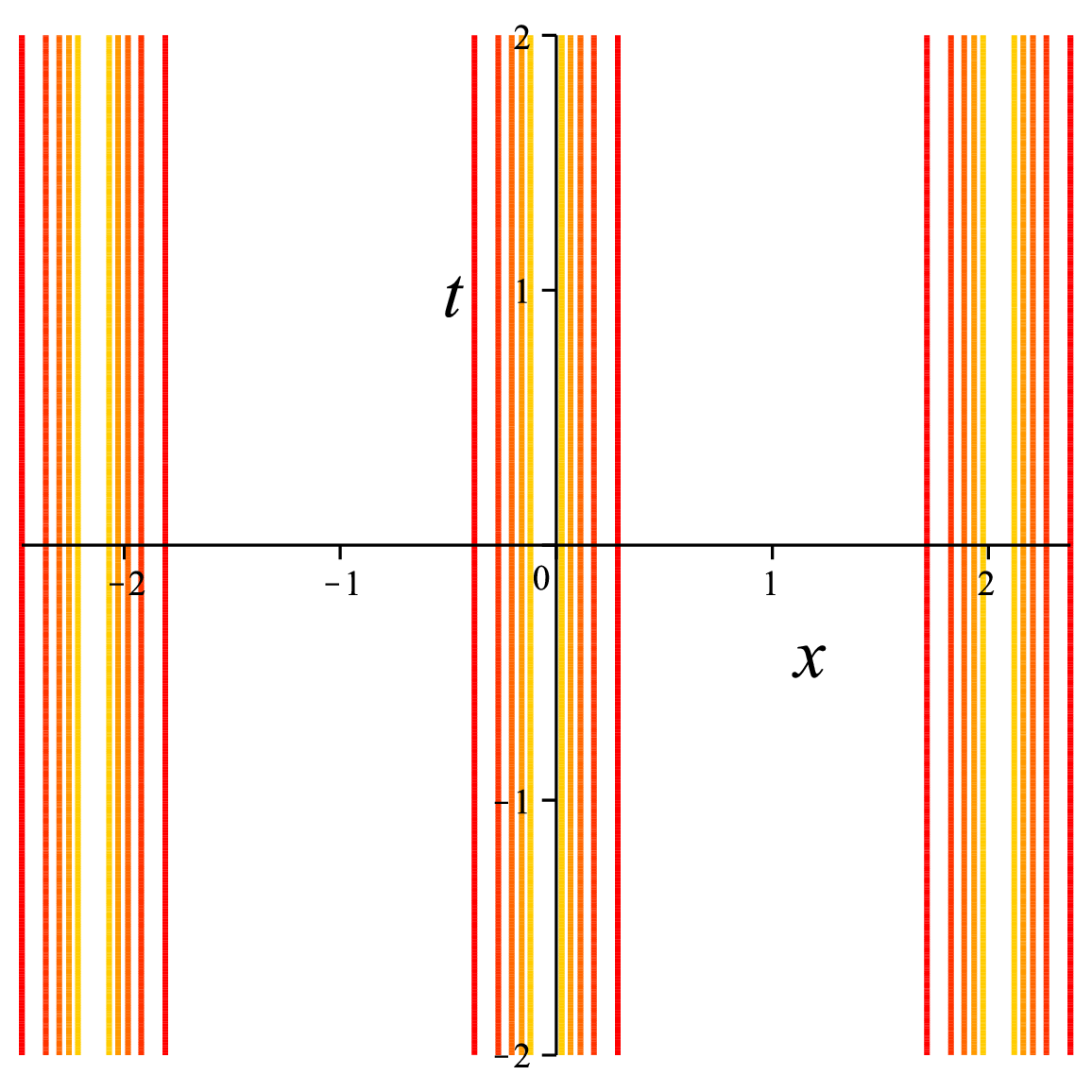}}
\caption{Periodic solution $|q|^2$ of the equation (\ref{twopex7}) (a) 3D graph, (b) contour plot.}
\end{figure}
\end{center}
\squeezeup

\noindent \textbf{Example 7.} Consider the equation (\ref{twopex8}) which is obtained from (\ref{4NLS-a})-(\ref{4NLS-d}) by the reduction formulas  $p(x,t)=q(x,t)$, $r(x,t)=\bar{q}(-x+x_0,-t+t_0)$, $s(x,t)=\bar{q}(-x+x_0,-t+t_0)$, and $c=\bar{c}$. Using these reductions with the solution (\ref{q-sol})-(\ref{s-sol}) yields
\begin{align}
&k_2=k_1,\,\,\, k_3=k_4=-\bar{k}_1,\,\,\, \omega_2=\omega_1,\,\,\,\omega_3=\omega_4=-\bar{\omega}_1,\nonumber\\
&e^{\delta_2}=e^{\delta_1},\,\,\, e^{\delta_3}=e^{\delta_4}=e^{\bar{\delta}_1+\bar{k}_1x_0+\bar{\omega}_1t_0}.
\end{align}
Therefore we obtain the following one-soliton solution of the equation (\ref{twopex8})
\begin{equation}\label{ex7sol}
q(x,t)=\frac{e^{k_1x+\omega_1t+\delta_1}}{1+\frac{2\sigma}{(k_1-\bar{k}_1)^2}e^{(k_1-\bar{k}_1)x+(\omega_1-\bar{\omega}_1)t+\delta_1+\bar{\delta}_1+\bar{k}_1x_0+\bar{\omega}_1t_0}}.
\end{equation}
Here $k_1-\bar{k}_1\neq 0$. Let $k_1=a+ib$ $(b\neq 0)$, $\delta_1=\alpha+i\beta$, then the denominator of the above solution, say $W(x,t)$ becomes
\begin{equation}
W(x,t)=1-\frac{\sigma}{2b^2}e^{\varphi}e^{i\phi},
\end{equation}
where
\begin{equation}
\varphi=2\alpha+ax_0+\Big(\frac{b^2-a^2}{c}\Big)t_0,\quad \phi=2bx-\frac{4ab}{c}t-bx_0+\frac{2ab}{c}t_0.
\end{equation}
If $W(x,t)=0$ then the solution (\ref{ex7sol}) becomes singular. This happens when $\frac{e^{\varphi}}{2b^2}=1$ and $e^{i\phi}=\sigma$.
Hence if $\sigma=1$, the solution becomes singular along the lines $2bx-\frac{4ab}{c}t=bx_0-\frac{2ab}{c}t_0+2n\pi$, and if $\sigma=-1$, through the lines
$2bx-\frac{4ab}{c}t=bx_0-\frac{2ab}{c}t_0+(2n+1)\pi$, $n\in \mathbb{Z}$ with the condition $2\alpha+ax_0+\Big(\frac{b^2-a^2}{c}\Big)t_0=\ln(2b^2)$ satisfied.
Therefore the solution (\ref{ex7sol}) is nonsingular if $2\alpha+ax_0+\Big(\frac{b^2-a^2}{c}\Big)t_0\neq \ln(2b^2)$.

Take the parameters of the solution (\ref{ex7sol}) as $k_1=1-i$, $c=x_0=t_0=\sigma=1$, and $\delta_1=i$ giving nonsingular solution
\begin{equation}\displaystyle
|q|^2=\frac{4e^{2x}}{e^2+4-4e\cos(2x-4t+1)}.
\end{equation}
The 3D and contour plot graphs of the above solution are given in the following Figure 7.
\begin{center}
\begin{figure}[h!]
\centering
\subfloat[]{\includegraphics[width=0.30\textwidth]{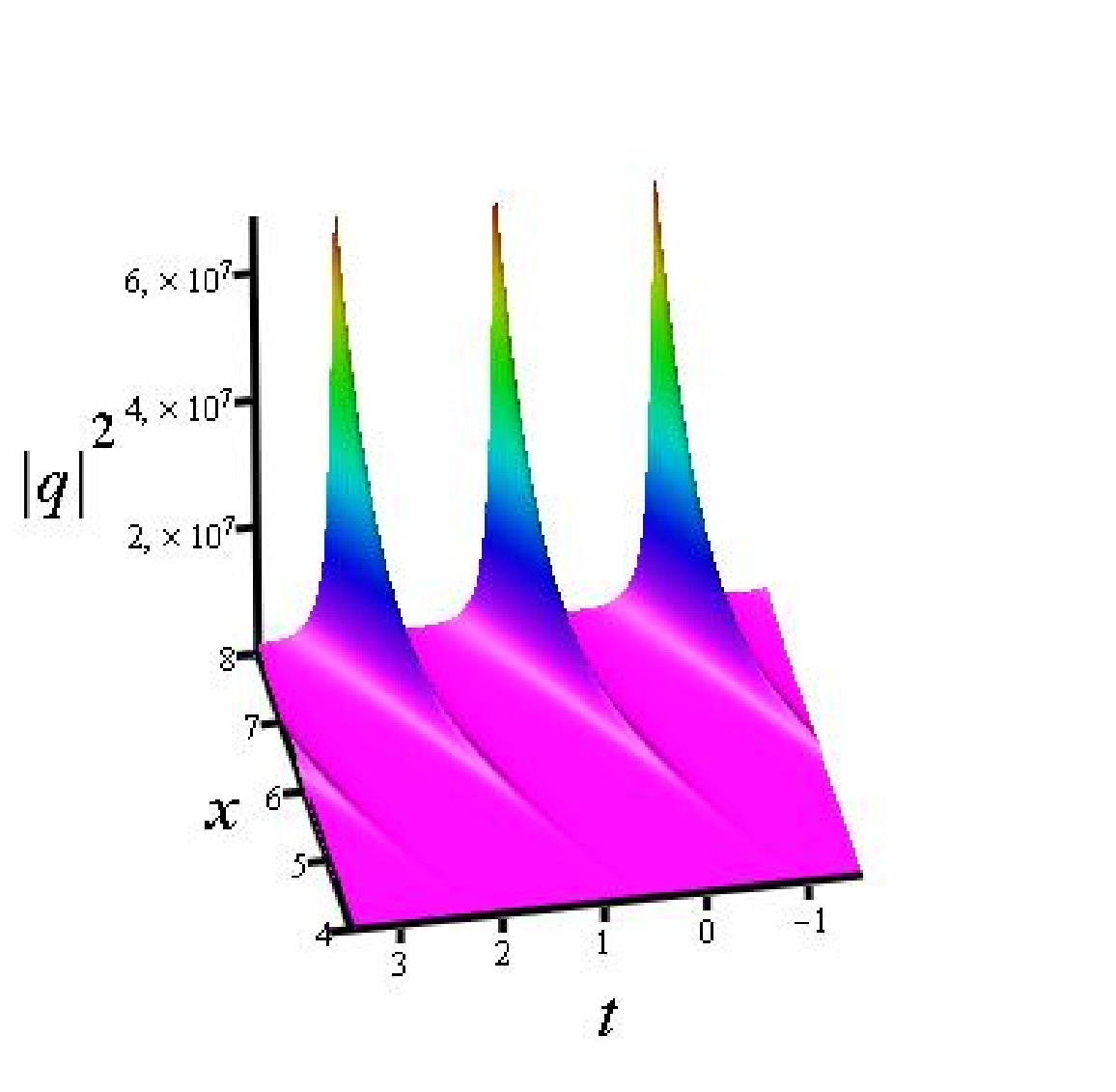}}\hspace{2cm}
\subfloat[]{\includegraphics[width=0.30\textwidth]{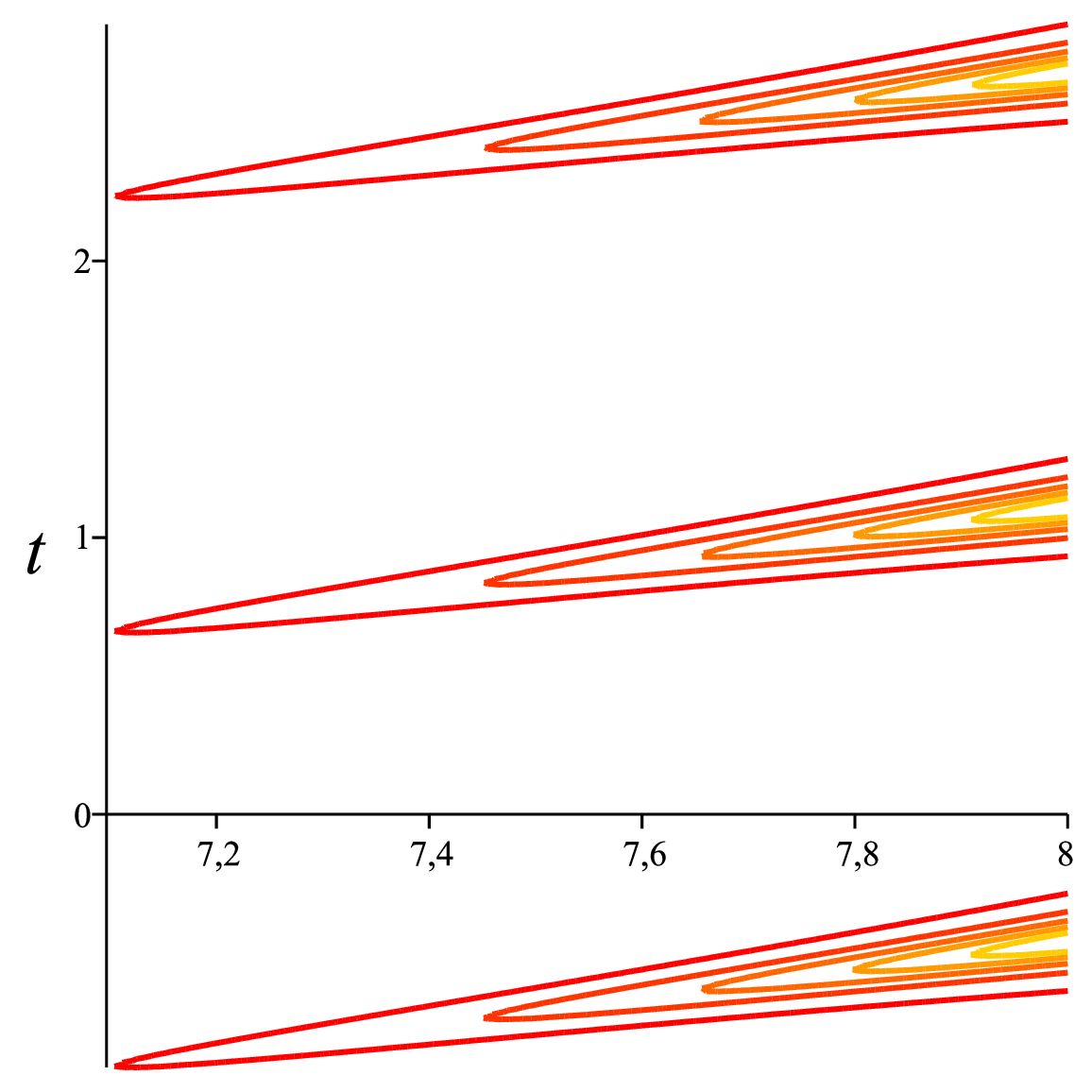}}
\caption{Periodic-type solution $|q|^2$ of the equation (\ref{twopex8}) (a) 3D graph, (b) contour plot.}
\end{figure}
\end{center}
\squeezeup

\noindent \textbf{Example 8.} Take the equation (\ref{twopex9}) which is obtained from the system (\ref{4NLS-a})-(\ref{4NLS-d}) by the reduction formulas
 i) $p(x,t)=\bar{q}(x,t)$, $r(x,t)=q(x,-t+t_0)$, $s(x,t)=\bar{q}(x,-t+t_0)$, and $c=\bar{c}$, or ii) $p(x,t)=\bar{q}(x,-t+t_0)$, $r(x,t)=q(x,-t+t_0)$, $s(x,t)=\bar{q}(x,t)$, and $c=-\bar{c}$. We use these reduction formulas with the solution (\ref{q-sol})-(\ref{s-sol}). For $i)$ we obtain the following constraints:
\begin{align}
&k_2=\bar{k}_1,\quad k_3=k_1,\quad k_4=\bar{k}_1,\quad \omega_2=\bar{\omega}_1,\quad \omega_3=-\omega_1,\quad \omega_4=-\bar{\omega}_1,\\
&e^{\delta_2}=e^{\bar{\delta}_1},\quad e^{\delta_3}=e^{\delta_1+\omega_1t_0},\quad e^{\delta_4}=e^{\bar{\delta}_1+\bar{\omega}_1t_0},
\end{align}
and for $ii)$ we have
\begin{align}
&k_2=\bar{k}_1,\quad k_3=k_1,\quad k_4=\bar{k}_1,\quad \omega_2=-\bar{\omega}_1,\quad \omega_3=-\omega_1,\quad \omega_4=\bar{\omega}_1,\\
&e^{\delta_2}=e^{\bar{\delta}_1+\bar{\omega}_1t_0},\quad e^{\delta_3}=e^{\delta_1+\omega_1t_0},\quad e^{\delta_4}=e^{\bar{\delta}_1}.
\end{align}
One-soliton solution of the equation (\ref{twopex9}) corresponding to both $i)$ and $ii)$ is
{\small \begin{equation}\label{ex8sol}
q(x,t)=\frac{e^{k_1x+\omega_1t+\delta_1}+\frac{\sigma (k_1-\bar{k}_1)}{4\bar{k}_1^2(k_1+\bar{k}_1)}e^{(k_1+2\bar{k}_1)x
+\omega_1t+\delta_1+2\bar{\delta}_1+\bar{\omega}_1t_0}  }
{1+\frac{\sigma}{4}[\frac{e^{2k_1x+2\delta_1+\omega_1t_0}}{k_1^2}+\frac{e^{2\bar{k}_1x+2\bar{\delta}_1+\bar{\omega}_1t_0}}{\bar{k}_1^2}  ]+\frac{(k_1-\bar{k}_1)^2e^{2(k_1+\bar{k}_1)x+2(\delta_1+\bar{\delta}_1)+(\omega_1+\bar{\omega}_1)t_0}}{16k_1^2\bar{k}_1^2(k_1+\bar{k}_1)^2}}.
\end{equation}}
Here $k_1+\bar{k}_1\neq 0$. The solution (\ref{ex8sol}) is singular when its denominator, say $W(x)$, vanishes. For singularity analysis let us consider the Case ii). Let $k_1=a+ib=re^{i\theta}$, where $r^2=a^2+b^2$, $\theta=\mathrm{arg} (k)$, $\delta_1=\alpha+i\beta$, and $c=i\xi$. The denominator
$W(x)$ becomes
\begin{equation}
W(x)=1+\frac{\sigma}{2r^2} \psi(x)\cos(\phi(x))-\frac{b^2}{16a^2r^4}\psi(x)^2,
\end{equation}
where
\begin{equation}
\psi(x)=e^{2ax-\frac{2ab}{\xi}t_0+2\alpha},\quad \phi(x)=2bx+\frac{(a^2-b^2)}{\xi}t_0+2\beta-2\theta.
\end{equation}
Clearly, if $a\neq 0$ and $b\neq 0$, $W(x)=0$ has at least one real root. Therefore let $b=0$ and take $\theta=0$. Hence the solution of the equation (\ref{twopex9})
\begin{equation}
q(x,t)=\frac{e^{ax+\frac{ia^2}{\xi}t+\alpha+i\beta}}{1+\frac{\sigma}{2a^2}e^{2ax+2\alpha}\cos(\frac{a^2}{\xi}t_0+2\beta)}
\end{equation}
is nonsingular if $\sigma\cos(\frac{a^2}{\xi}t_0+2\beta)\geq 0$.

Let us consider the following particular example of a solution. Take $\sigma=1$, $c=i$, $k_1=2$, $\delta_1=1+2i$, $t_0=4$. Then the real-valued solution $|q|^2$ of the equation (\ref{twopex9}) becomes
\begin{equation}
|q|^2=\frac{e^{4x+2}}{(1+\frac{1}{8}e^{4x+2}\cos(20))^2}.
\end{equation}
The 3D and contour plot graphs of the above solution $|q|^2$ are given in the following Figure 8.
\begin{center}
\begin{figure}[h!]
\centering
\subfloat[]{\includegraphics[width=0.30\textwidth]{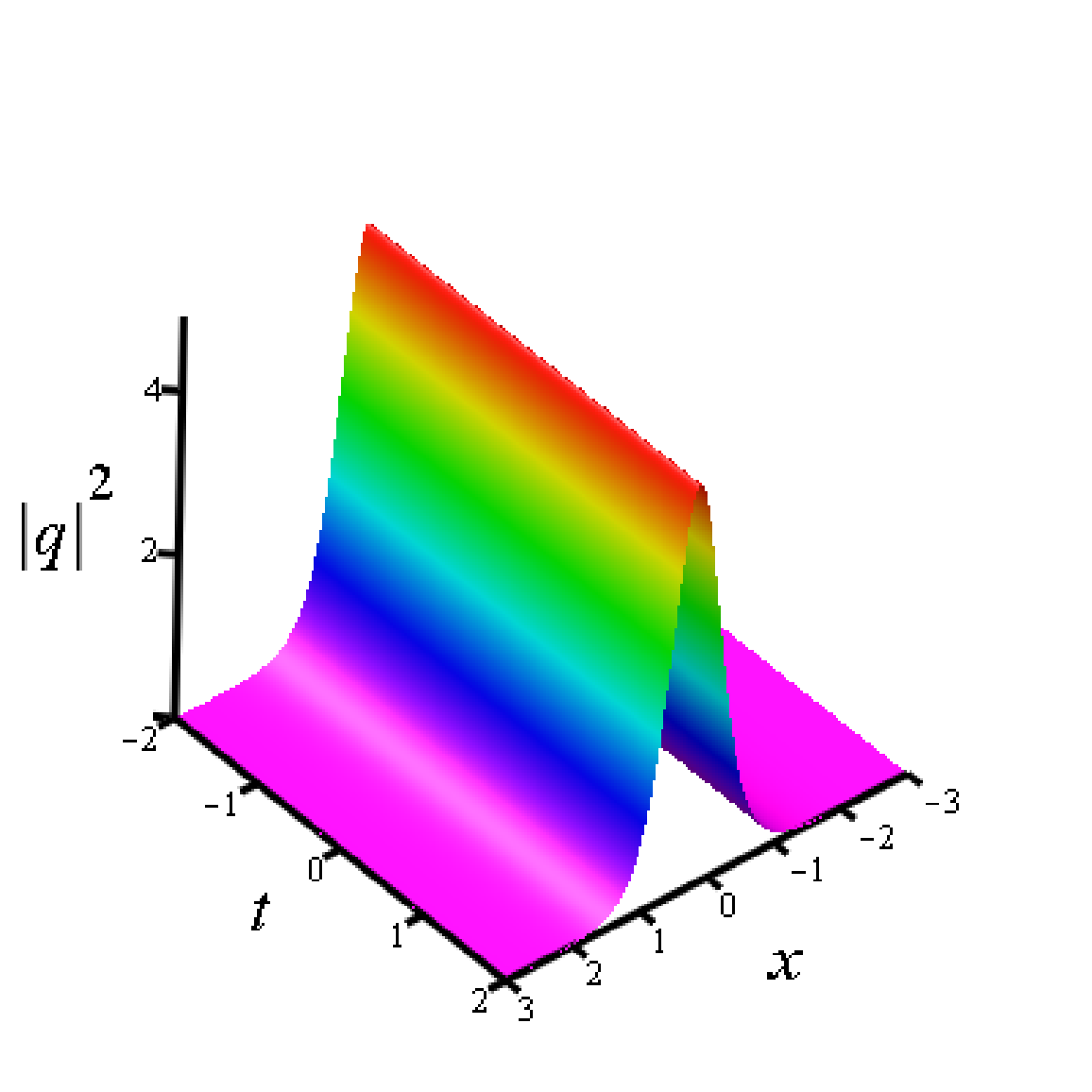}}\hspace{2cm}
\subfloat[]{\includegraphics[width=0.30\textwidth]{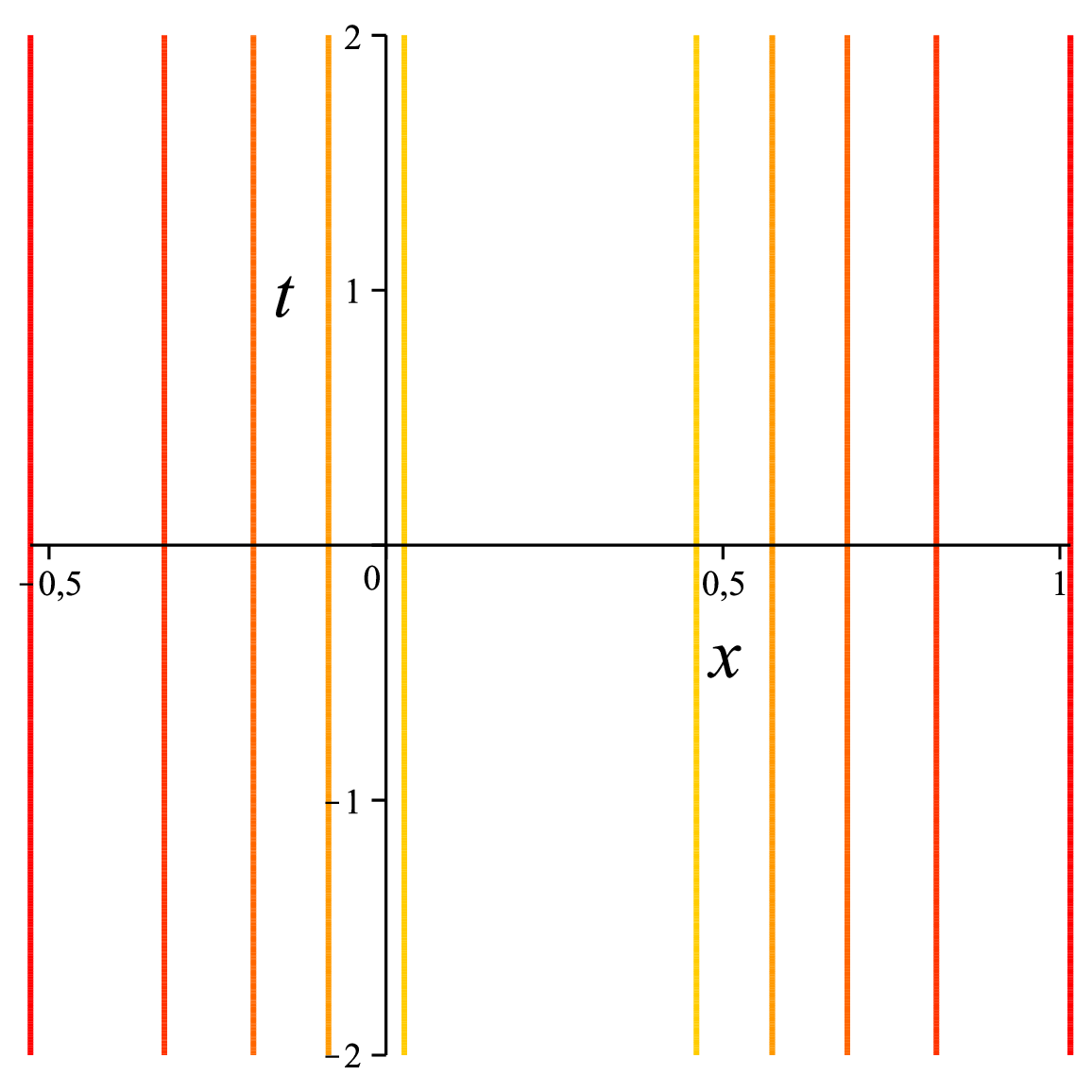}}
\caption{Bell-type soliton solution $|q|^2$ of the equation (\ref{twopex9}) (a) 3D graph, (b) contour plot.}
\end{figure}
\end{center}
\squeezeup

\noindent \textbf{Example 9.} Consider the equation (\ref{twopex13}) which is obtained from the system (\ref{4NLS-a})-(\ref{4NLS-d}) by the reduction formulas
$p(x,t)=\bar{q}(x,t)$, $r(x,t)=\bar{q}(x,-t+t_0)$, $s(x,t)=q(x,-t+t_0)$, and $c=\bar{c}$. We use these reduction formulas with the solution (\ref{q-sol})-(\ref{s-sol}). We get
\begin{align}
&k_2=k_3=\bar{k}_1,\quad k_4=k_1,\quad \omega_2=\bar{\omega}_1,\quad \omega_3=-\bar{\omega}_1,\quad \omega_4=-\omega_1,\\
&e^{\delta_2}=e^{\bar{\delta}_1},\quad e^{\delta_3}=e^{\bar{\delta}_1+\bar{\omega}_1t_0},\quad e^{\delta_4}=e^{\delta_1+\omega_1t_0}.
\end{align}
Therefore one-soliton solution of the two-place shifted time reversal equation (\ref{twopex13}) is
{\small\begin{equation}\label{ex9sol}
q(x,t)=\frac{e^{k_1x+\omega_1t+\delta_1}+\frac{\sigma(k_1-\bar{k}_1)e^{(2k_1+\bar{k}_1)x+\bar{\omega}_1t+2\delta_1+\bar{\delta}_1
+\omega_1t_0}}{2k_1(k_1+\bar{k}_1)^2}}{1+\frac{\sigma e^{(k_1+\bar{k}_1)x+\delta_1+\bar{\delta}_1}}
{(k_1+\bar{k}_1)^2}[e^{(\omega_1-\bar{\omega}_1)t+\bar{\omega}_1t_0}+e^{-(\omega_1-\bar{\omega}_1)t+\omega_1t_0 }]-\frac{(k_1-\bar{k}_1)^2e^{2(k_1+\bar{k}_1)x+2(\delta_1+\bar{\delta}_1)+(\omega_1+\bar{\omega}_1)t_0  } }{4k_1\bar{k}_1(k_1+\bar{k}_1)^4}}.
\end{equation}}
Here $k_1+\bar{k}_1\neq 0$. Let $k_1=a+ib$, $\delta_1=\alpha+i\beta$. Then the denominator, say $W(x,t)$, of the solution (\ref{ex9sol}) becomes
{\small \begin{equation}
W(x,t)=1+\sigma Y(x)\cos\Big(\frac{4ab}{c}t-\frac{2ab}{c}t_0 \Big)+\frac{b^2}{4(a^2+b^2)}Y^2(x),\quad Y(x)=\frac{e^{2ax+2\alpha+(\frac{b^2-a^2}{c})t_0}}{2a^2}>0.
\end{equation}}
If $b\neq 0$, then we can find some $(x,t)$ so that $W(x,t)=0$. If $b=0$, then the denominator becomes a function of $x$; $W(x)=1+\sigma Y(x)$.
If $\sigma=1$, then $W(x)>0$. If $\sigma=-1$, we can find some $x\in \mathbb{R}$ so that $W(x)=0$. Hence  if $a\neq 0$, $b=0$, and $\sigma=1$, the solution (\ref{ex9sol}) is nonsingular.

Consider the following particular example of a solution. Take $\sigma=1$, $c=k_1=2$, $\delta_1=1+i$, $t_0=1$. Therefore the real-valued solution $|q|^2$ of the equation (\ref{twopex13}) becomes
\begin{equation}
|q|^2=\frac{64e^{4x-4t+2}}{(e^{4x}+8)^2}.
\end{equation}
The 3D and contour plot graphs of the above solution $|q|^2$ are given in the following Figure 9.
\begin{center}
\begin{figure}[h!]
\centering
\subfloat[]{\includegraphics[width=0.29\textwidth]{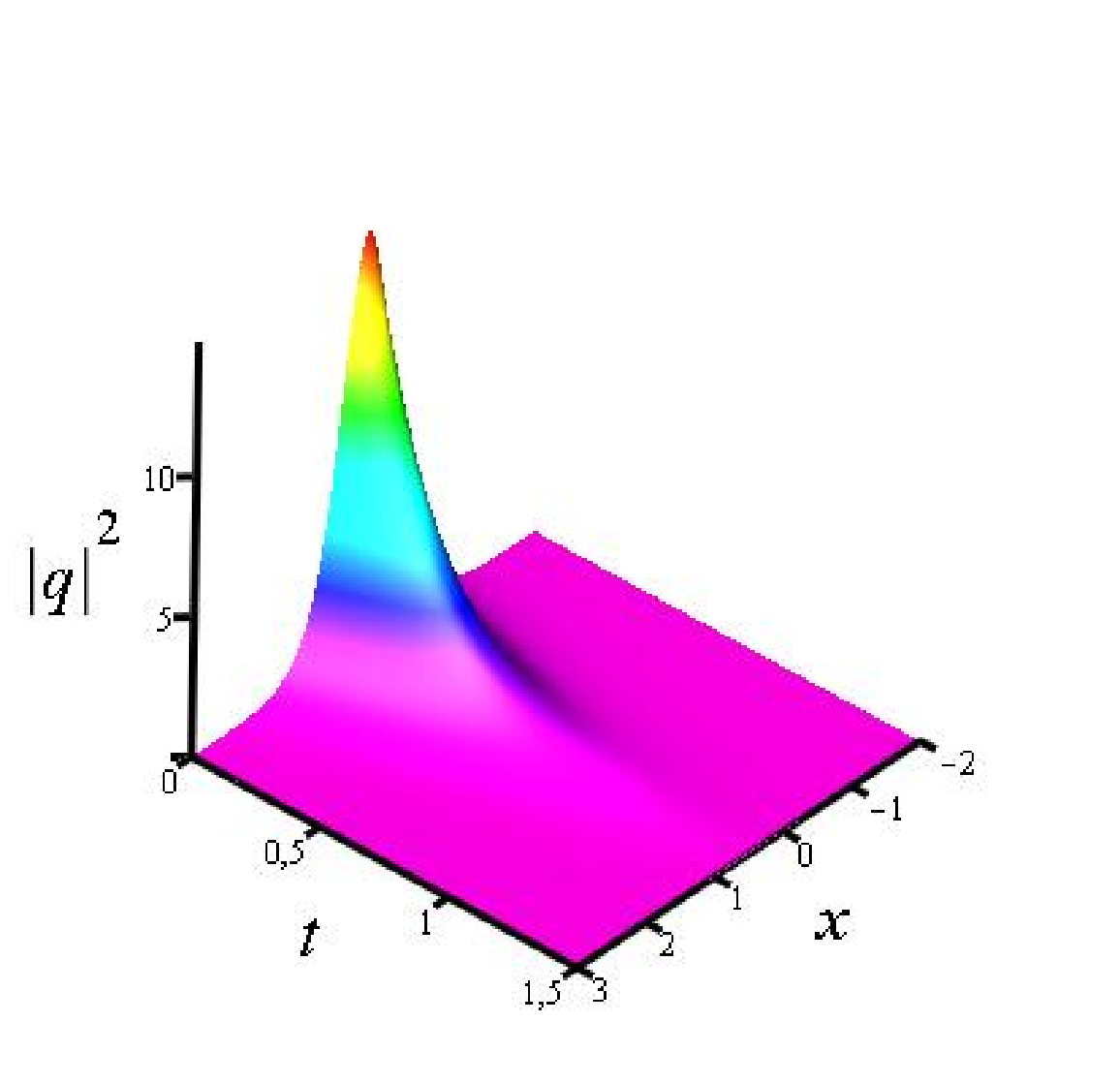}}\hspace{2cm}
\subfloat[]{\includegraphics[width=0.29\textwidth]{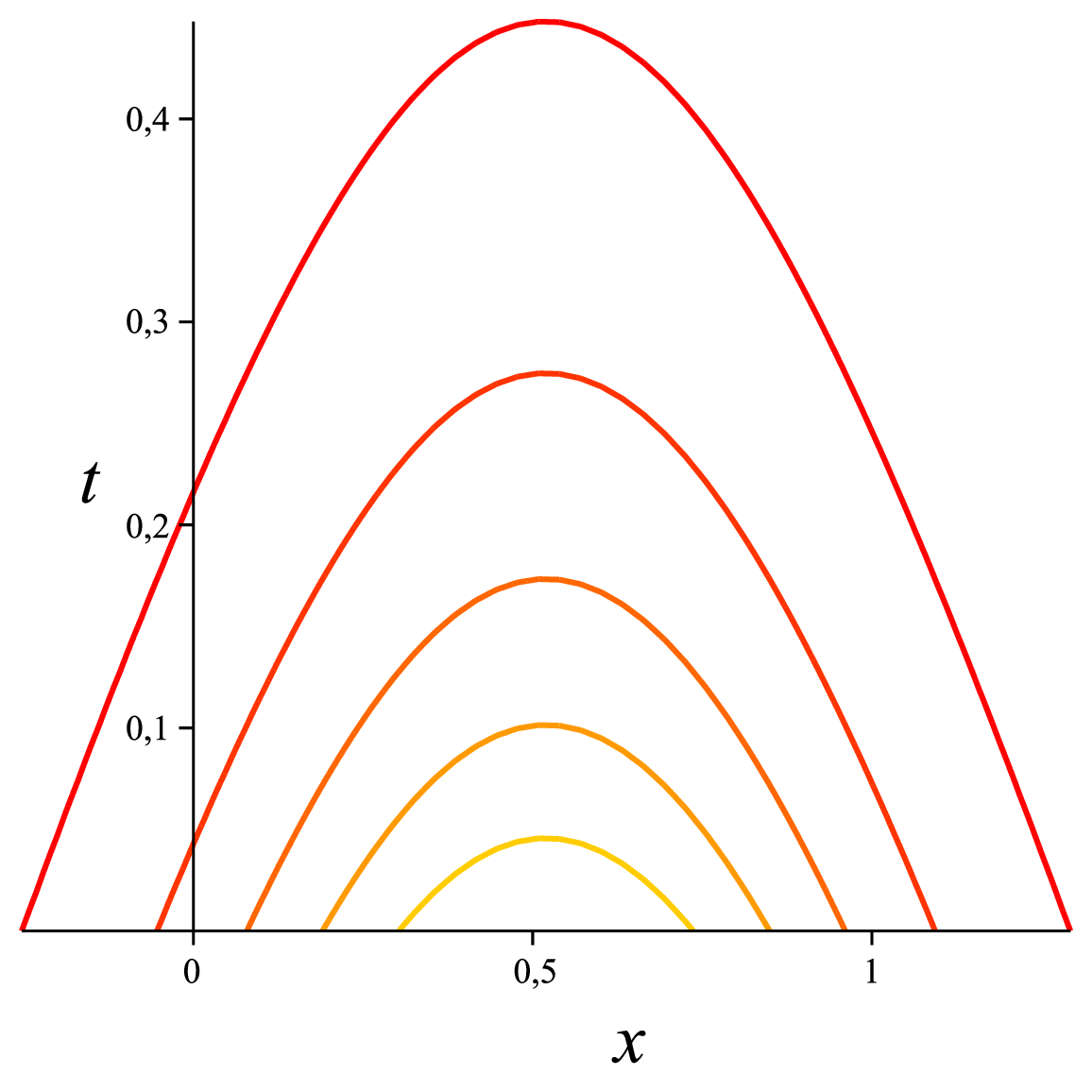}}
\caption{Asymptotically decaying solution $|q|^2$ of the equation (\ref{twopex13}) (a) 3D graph, (b) contour plot.}
\end{figure}
\end{center}
\squeezeup
\newpage
\noindent \textbf{Remark.} We obtain trivial solutions for the two-place shifted nonlocal equations (\ref{twopex5}), (\ref{twopex10})-Case ii, and (\ref{twopex11}) by our method. The constraints corresponding to the reductions leading to equations (\ref{twopex10})-Case i and (\ref{twopex12})  give $k_4=-k_1$ which makes the denominator in the general one-soliton formula vanish. Hence these cases do not yield nontrivial solutions by the present one-soliton ansatz.

\section{Solutions of four-place shifted nonlocal equations}

In this part we shall give examples of soliton solutions of four-place shifted nonlocal equations.

\noindent \textbf{Example 10.} Take the equation (\ref{four-ex3}) which is obtained from (\ref{4NLS-a})-(\ref{4NLS-d}) by the reduction formulas
$p(x,t)=q(-x+x_0,t)$, $r(x,t)=\bar{q}(-x+x_0,-t+t_0)$, $s(x,t)=\bar{q}(x,-t+t_0)$, and $c=\bar{c}$. Using these reductions with the solution (\ref{q-sol})-(\ref{s-sol}) we  obtain
\begin{align}
&k_2=-k_1,\quad k_3=-\bar{k}_1,\quad k_4=\bar{k}_1,\quad \omega_2=\omega_1,\quad \omega_3=-\bar{\omega}_1,\quad \omega_4=-\bar{\omega}_1,\\
&e^{\delta_2}=e^{\delta_1+k_1x_0},\quad e^{\delta_3}=e^{\bar{\delta}_1+\bar{k}_1x_0+\bar{\omega}_1t_0},\quad e^{\delta_4}=e^{\bar{\delta}_1+\bar{\omega}_1t_0}.
\end{align}
So one-soliton solution of the four-place shifted nonlocal equation (\ref{four-ex3}) is
{\small\begin{align}\label{ex10sol}
&q(x,t)=\Bigg(e^{k_1x+\omega_1t+\delta_1}+\frac{2\sigma k_1e^{\bar{k}_1x+(2\omega_1-\bar{\omega}_1)t+2\delta_1+\bar{\delta}_1
+k_1x_0+\bar{\omega}_1t_0}}{(k_1+\bar{k}_1)(k_1-\bar{k}_1)^2}\Bigg)\Bigg/
\Bigg(1+\frac{\sigma e^{(\omega_1-\bar{\omega}_1)t+\delta_1+\bar{\delta}_1+\bar{\omega}_1t_0}}
{(k_1-\bar{k}_1)^2}\nonumber\\
&\times [e^{(k_1-\bar{k}_1)x+\bar{k}_1x_0}+e^{-(k_1-\bar{k}_1)x+k_1x_0 }]+\frac{4k_1\bar{k}_1e^{2(\omega_1-\bar{\omega}_1)t+2(\delta_1+\bar{\delta}_1)+(k_1+\bar{k}_1)x_0+2\bar{\omega}_1t_0 }}{(k_1+\bar{k}_1)^2(k_1-\bar{k}_1)^4}\Bigg).
\end{align}}
Here $k_1+\bar{k}_1\neq 0$ and $k_1-\bar{k}_1\neq 0$. The solution (\ref{ex10sol}) is singular when its denominator, say $W(x,t)$, vanishes. Let $k_1=a+ib$ and $\delta_1=\alpha+i\beta$. The denominator
$W(x,t)$ becomes
\begin{align}
W(x,t)&=1-\sigma\rho e^{i\psi(t)}\cos(2bx-bx_0)+\rho^2\mu e^{2i\psi(t)},\nonumber\\
&=\mathrm{Re}(W(x,t))+i\mathrm{Im}(W(x,t))
\end{align}
where
\begin{align}
\mathrm{Re}(W(x,t))=1-\sigma \rho\cos(\psi(t))\cos(2bx-bx_0)+\rho^2\mu\cos(2\psi(t)),\\
\mathrm{Im}(W(x,t))=\sin(\psi(t))[-\sigma\rho\cos(2bx-bx_0)+2\rho^2\mu\cos(\psi(t))],
\end{align}
for
\begin{equation}
\psi(t)=-\frac{4ab}{c}t+\frac{2ab}{c}t_0,\quad \rho=\frac{e^{2\alpha+\frac{(b^2-a^2)}{c}t_0+ax_0}}{2b^2},\quad \mu=\frac{a^2+b^2}{4a^2}.
\end{equation}
We now determine the cases giving $\mathrm{Im}(W(x,t))=0$.

\noindent Case i. Let $\sin(\psi(t))=0$, i.e., $\psi(t)=n\pi$, $n\in \mathbb{Z}$. In this case the denominator $W(x,t)$ takes the form
\begin{equation}
W(x,t)=1-\sigma \rho (-1)^n\cos(2bx-bx_0)+\rho^2\mu.
\end{equation}
This function vanishes when $\cos(\theta(x))=(-1)^n\sigma\frac{1+\rho^2\mu}{\rho}$, $\theta(x)=2bx-bx_0$. Note that
\begin{equation}
\frac{1+\rho^2\mu}{\rho}=\rho\mu+\frac{1}{\rho}\geq 2\sqrt{\mu}
\end{equation}
yielding $|\cos(\theta(x))|>1$ which is a contradiction. Hence $W(x,t)\neq 0$ that is the solution (\ref{ex10sol}) is nonsingular in this case.

\noindent Case ii. Let $\sin(\psi(t))\neq 0$. In this case to have $\mathrm{Im}(W(x,t))=0$
\begin{equation}
\cos(\theta(x))=2\mu \rho \sigma\cos(\psi(t)).
\end{equation}
Use the above equality in $\mathrm{Re}(W(x,t))$. We have
\begin{equation}
\mathrm{Re}(W(x,t))=1-\mu\rho^2,
\end{equation}
which is zero if $\mu\rho^2=1$. Hence if $\sin(\psi(t))\neq 0$ and $e^{4\alpha+2\frac{(b^2-a^2)}{c}t_0+2ax_0  }=\frac{16a^2b^4}{a^2+b^2}$, the solution
(\ref{ex10sol}) is singular. For nonsingularity, we must have  $\mu\rho^2\neq 1$ if $\sin(\psi(t))\neq 0$.

Let us take particular values for the parameters so that the solution (\ref{ex10sol}) is nonsingular. Choose $c=1$, $k_1=\frac{1}{2}+i\frac{1}{2}$, $\delta_1=0$, $\sigma=-1$, and
$x_0=t_0=1$. Hence the solution (\ref{ex10sol}) becomes
\begin{equation}
q(x,t)=\frac{e^{-\frac{i}{2}t}\left(e^{\frac{x}{2}+\frac{i}{2}x}+e^{\frac{x}{2}+\frac{1}{2}+i-\frac{i}{2}x-it}+i e^{\frac{x}{2}+\frac{1}{2}+i-\frac{i}{2}x-it}
\right)}{1+e^{-it+ix+\frac{1}{2}}+e^{-it+\frac{1}{2}+i-ix}+2e^{-2it+1+i}}.
\end{equation}
The expression for $|q|^2$ is lengthy and is therefore omitted. We give the 3D and contour plot graphs of the solution $|q|^2$ in the following Figure 10.
\begin{center}
\begin{figure}[h!]
\centering
\subfloat[]{\includegraphics[width=0.30\textwidth]{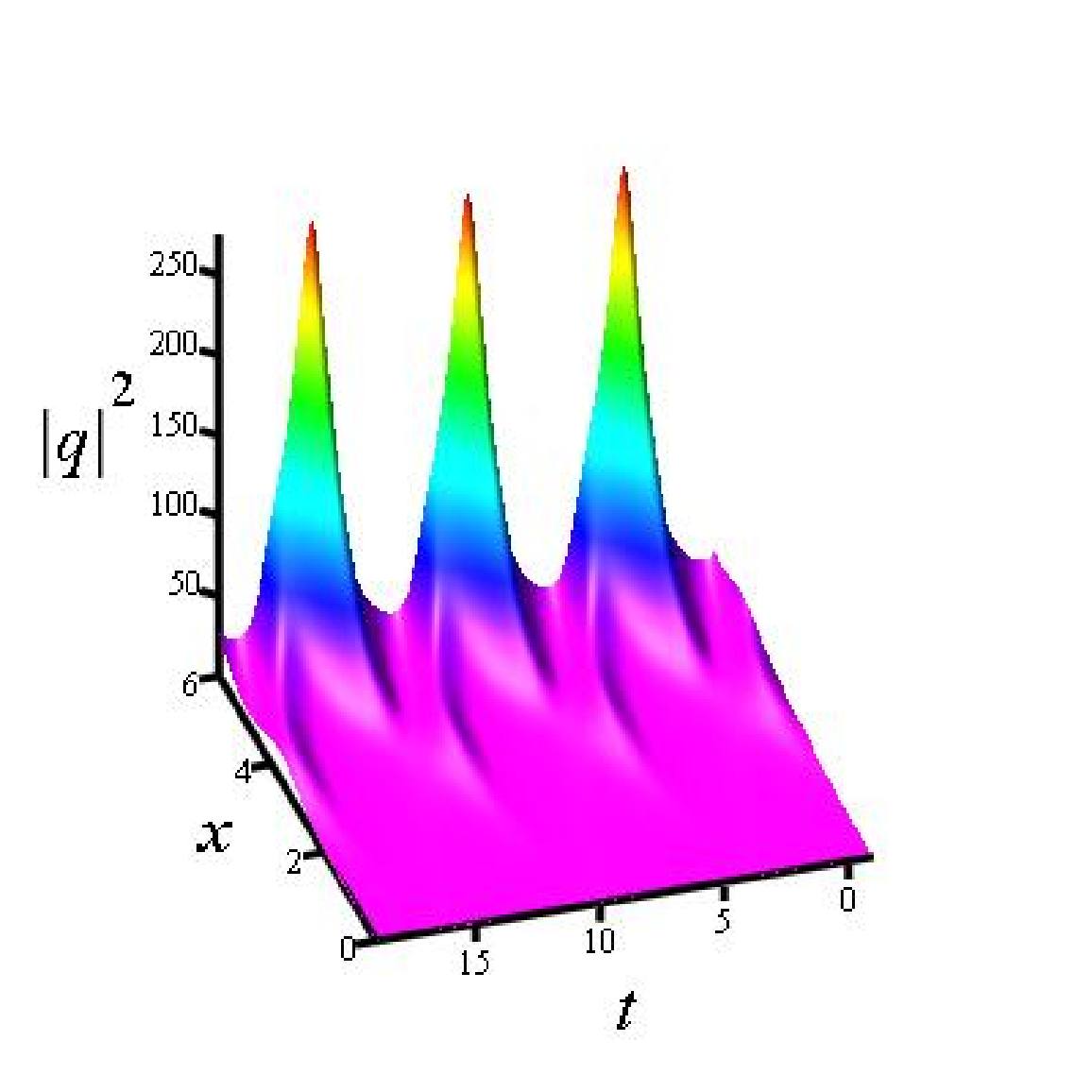}}\hspace{2cm}
\subfloat[]{\includegraphics[width=0.30\textwidth]{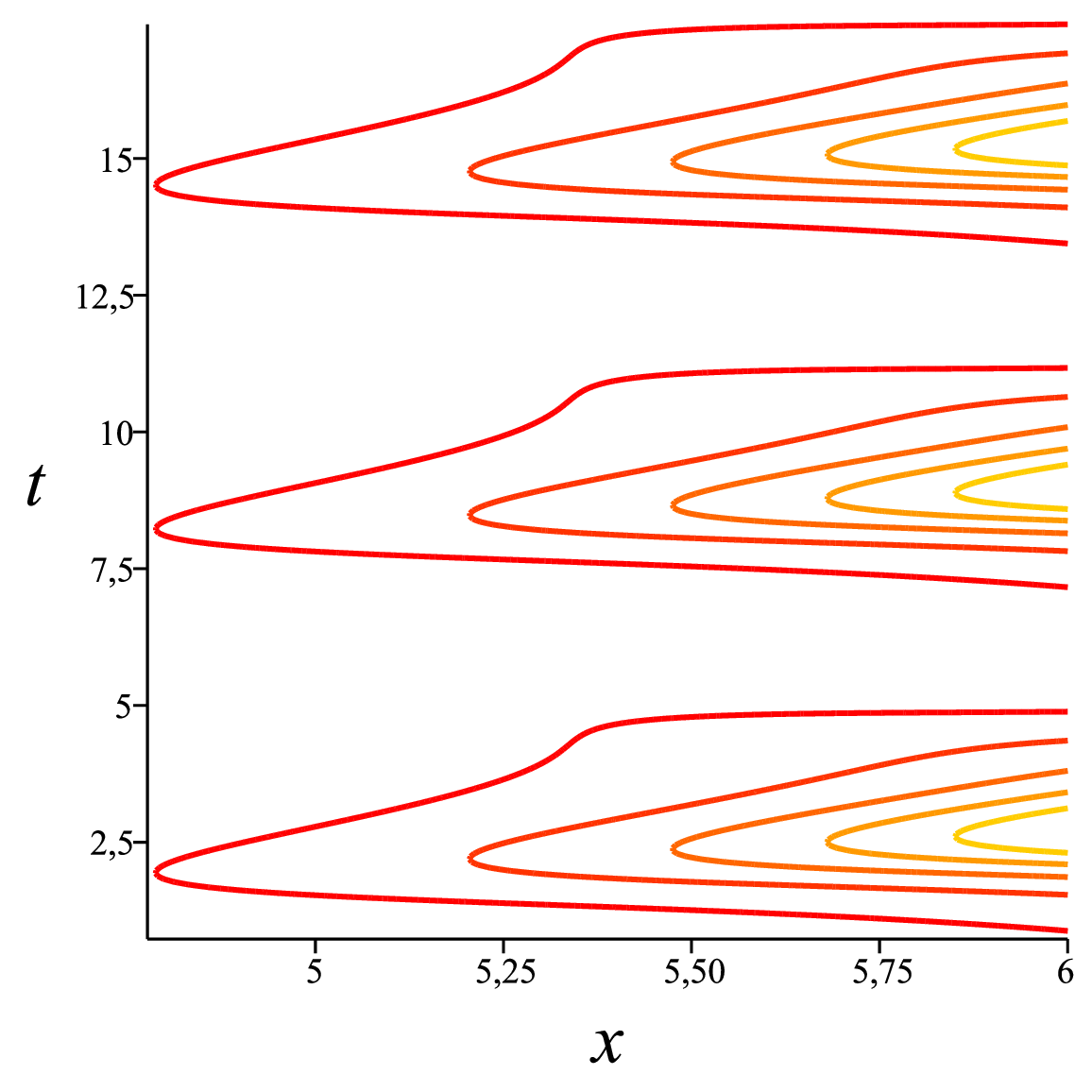}}
\caption{Periodic-type solution $|q|^2$ of the equation (\ref{four-ex3}) (a) 3D graph, (b) contour plot.}
\end{figure}
\end{center}
\squeezeup

\noindent \textbf{Example 11.} Consider the equation (\ref{four-ex4}) which is derived from (\ref{4NLS-a})-(\ref{4NLS-d}) by the reduction formulas
$p(x,t)=q(-x+x_0,t)$, $r(x,t)=\bar{q}(x,-t+t_0)$, $s(x,t)=\bar{q}(-x+x_0,-t+t_0)$, and $c=\bar{c}$. By using these reduction formulas with the solution (\ref{q-sol})-(\ref{s-sol}), we obtain
\begin{align}
&k_2=-k_1,\quad k_3=\bar{k}_1,\quad k_4=-\bar{k}_1,\quad \omega_2=\omega_1,\quad \omega_3=-\bar{\omega}_1,\quad \omega_4=-\bar{\omega}_1,\\
&e^{\delta_2}=e^{\delta_1+k_1x_0},\quad e^{\delta_3}=e^{\bar{\delta}_1+\bar{\omega}_1t_0},\quad e^{\delta_4}=e^{\bar{\delta}_1+\bar{k}_1x_0+\bar{\omega}_1t_0}.
\end{align}
Hence one-soliton solution of the four-place shifted nonlocal equation (\ref{four-ex4}) is
{\small \begin{align}\label{ex11sol}
&q(x,t)=\Bigg(e^{k_1x+\omega_1t+\delta_1}+\frac{4\sigma k_1e^{-\bar{k}_1x+(2\omega_1-\bar{\omega}_1)t+2\delta_1+\bar{\delta}_1
+(k_1+\bar{k}_1)x_0+\bar{\omega}_1t_0}}{(k_1-\bar{k}_1)(k_1+\bar{k}_1)^2}\Bigg)\Bigg/\Bigg(1+\frac{2\sigma e^{(\omega_1-\bar{\omega}_1)t+\delta_1+\bar{\delta}_1+\bar{\omega}_1t_0}}
{(k_1+\bar{k}_1)^2}\nonumber\\
&\times [e^{(k_1+\bar{k}_1)x}+e^{-(k_1+\bar{k}_1)x+(k_1+\bar{k}_1)x_0 }]-\frac{16k_1\bar{k}_1e^{2(\omega_1-\bar{\omega}_1)t+2(\delta_1+\bar{\delta}_1)+(k_1+\bar{k}_1)x_0+2\bar{\omega}_1t_0 }}{(k_1-\bar{k}_1)^2(k_1+\bar{k}_1)^4}
\Bigg).
\end{align}}
Here $k_1+\bar{k}_1\neq 0$ and $k_1-\bar{k}_1\neq 0$. Let $k_1=a+ib$ and $\delta_1=\alpha+i\beta$. Here $c \in \mathbb{R}$ and $a\neq 0$, $b\neq 0$. Consider the denominator of the solution (\ref{ex11sol}), say $W(x,t)$, for checking singularity of the solution. We have
\begin{equation}\label{ex11denom}
W(x,t)=1+Y(x)e^{i\phi(t)}+B_0e^{2i\phi(t)},
\end{equation}
where
\begin{align}\displaystyle
& Y(x)=\frac{\sigma A_0}{a^2}\cosh(2ax-ax_0),\quad B_0=\frac{a^2+b^2}{4b^2a^4}A_0^2,\nonumber\\
& A_0=e^{2\alpha+\Big(\frac{b^2-a^2}{c}\Big)t_0+ax_0},\quad \phi(t)=-\frac{4ab}{c}t+\frac{2ab}{c}t_0.
\end{align}
Clearly, $A_0>0$ and $B_0>0$. The solution (\ref{ex11sol}) is singular when $W(x,t)=0$, which is equivalent to $\tilde{W}(x,t)=e^{-i\phi(t)}+Y(x)+B_0e^{i\phi(t)}=0$. We have
\begin{align}
&\mathrm{Re}(\tilde{W}(x,t))=(1+B_0)\cos(\phi(t))+Y(x)=0,\nonumber\\
&\mathrm{Im}(\tilde{W}(x,t))=(B_0-1)\sin(\phi(t))=0.
\end{align}
From $\mathrm{Im}(\tilde{W}(x,t))=0$ we have two possibilities.

\noindent Case i. $B_0\neq 1$ so $\sin(\phi(t))=0$, i.e., $\phi(t)=n\pi$, $n\in \mathbb{Z}$. Hence we have $e^{i\phi(t)}=\pm 1$. For $\sigma=-1$ we have $Y(x)<0$.
If we choose $e^{i\phi(t)}=1$, then we check whether the equality $W(x,t)=1+Y(x)+B_0=0$ is solvable for real $(x,t)$. Since
\begin{equation}
\frac{A_0}{a^2}\cosh(2ax-ax_0)=1+B_0
\end{equation}
is satisfied for some $x\in \mathbb{R}$ due to the fact that $\frac{a^2(1+B_0)}{A_0}>1$, the solution (\ref{ex11sol}) is singular at some $(x,t)$. We get the same
result for $\sigma=1$ and $e^{i\phi(t)}=-1$.

\noindent Case ii. Let $B_0=1$. This makes $\mathrm{Re}(\tilde{W}(x,t))=2\cos(\phi(t))+Y(x)=0$ that is $\cos(\phi(t))=-\frac{Y(x)}{2}$. A real solution to this equality exists provided that $|Y(x)|\leq 2$. This condition is satisfied for some $x\in \mathbb{R}$. Hence in this special case, there exist real $(x,t)$ making the denominator $W(x,t)=0$.

Therefore for both $\sigma=\pm 1$ we conclude that the solution (\ref{ex11sol}) develops singularities for some real $(x,t)$. Hence we observe that we cannot derive nonsingular solutions for the equation (\ref{four-ex4}) by our solution method.
\squeezeup

\noindent \textbf{Example 12.} Take the equation (\ref{four-ex5}) which is obtained from (\ref{4NLS-a})-(\ref{4NLS-d}) by the reduction formulas
$p(x,t)=\bar{q}(-x+x_0,-t+t_0)$, $r(x,t)=q(x,-t+t_0)$, $s(x,t)=\bar{q}(-x+x_0,t)$, and $c=-\bar{c}$. By using these reductions with the solution (\ref{q-sol})-(\ref{s-sol}) we get the constraints:
\begin{align}
&k_2=-\bar{k}_1,\quad k_3=k_1,\quad k_4=-\bar{k}_1,\quad \omega_2=-\bar{\omega}_1,\quad \omega_3=-\omega_1,\quad \omega_4=\bar{\omega}_1,\\
&e^{\delta_2}=e^{\bar{\delta}_1+\bar{k}_1x_0+\bar{\omega}_1t_0},\quad e^{\delta_3}=e^{\delta_1+\omega_1t_0},\quad e^{\delta_4}=e^{\bar{\delta}_1+\bar{k}_1x_0}.
\end{align}
So one-soliton solution of the four-place shifted nonlocal equation (\ref{four-ex5}) is
{\small \begin{equation}\label{ex12sol}
q(x,t)=\frac{e^{k_1x+\omega_1t+\delta_1}+\frac{\sigma (k_1+\bar{k}_1)e^{(k_1-2\bar{k}_1)x+\omega_1t+\delta_1+2\bar{\delta}_1
+2\bar{k}_1x_0+\bar{\omega}_1t_0}}{2\bar{k}_1^2(k_1-\bar{k}_1)}}{1+\frac{\sigma }
{2}[\frac{e^{2k_1x+2\delta_1+\omega_1t_0}}{k_1^2}+\frac{e^{-2\bar{k}_1x+2\bar{\delta}_1+2\bar{k}_1x_0+\bar{\omega}_1t_0}}{\bar{k}_1^2} ]+\frac{(k_1+\bar{k}_1)^2e^{2(k_1-\bar{k}_1)x+2(\delta_1+\bar{\delta}_1)+2\bar{k}_1x_0+(\omega_1+\bar{\omega}_1)t_0 }}{4k_1^2\bar{k}_1^2(k_1-\bar{k}_1)^2}
}.
\end{equation}}
Here $k_1-\bar{k}_1\neq 0$. Let $k_1=ib$ for $b\neq 0$, $c=i\xi$, and $\delta_1=\alpha+i\beta$. Then the denominator of the solution (\ref{ex12sol}), say $W(x)$, becomes
\begin{equation}
W(x)=1-\frac{\sigma}{b^2}e^{2\alpha}e^{i(2bx-bx_0)}\cos\Big( 2\beta-\frac{b^2}{\xi}t_0+bx_0\Big).
\end{equation}
For nonsingularity of the solution we need to have $W(x)\neq 0$. This happens when $\frac{e^{2\alpha}}{b^2}|\sigma \cos\Big( 2\beta-\frac{b^2}{\xi}t_0+bx_0\Big)  |\neq 1$. Let us give a particular example of a nonsingular solution. Choose $c=k_1=i$, $\delta_1=-1$, $\sigma=1$, and
$x_0=1$, $t_0=2$. Therefore the solution (\ref{ex12sol}) becomes
\begin{equation}
|q|^2=\frac{2e^{-2}  }{2+e^{-4}(1+\cos(2))-2e^{-2}[\cos(2x)+\cos(2x-2)]}.
\end{equation}
We give the 3D and contour plot graphs of the solution $|q|^2$ in the following Figure 11.
\begin{center}
\begin{figure}[h!]
\centering
\subfloat[]{\includegraphics[width=0.30\textwidth]{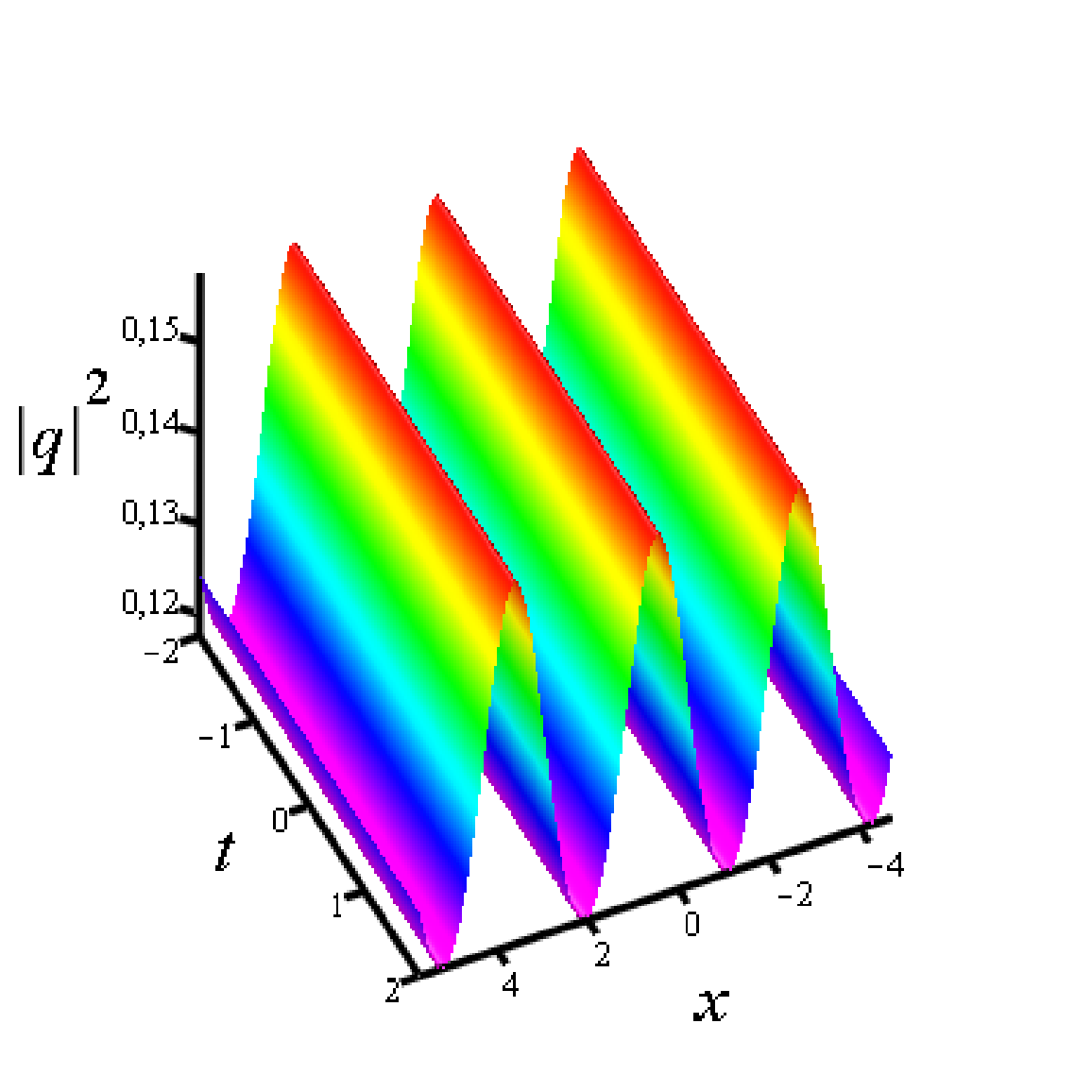}}\hspace{2cm}
\subfloat[]{\includegraphics[width=0.30\textwidth]{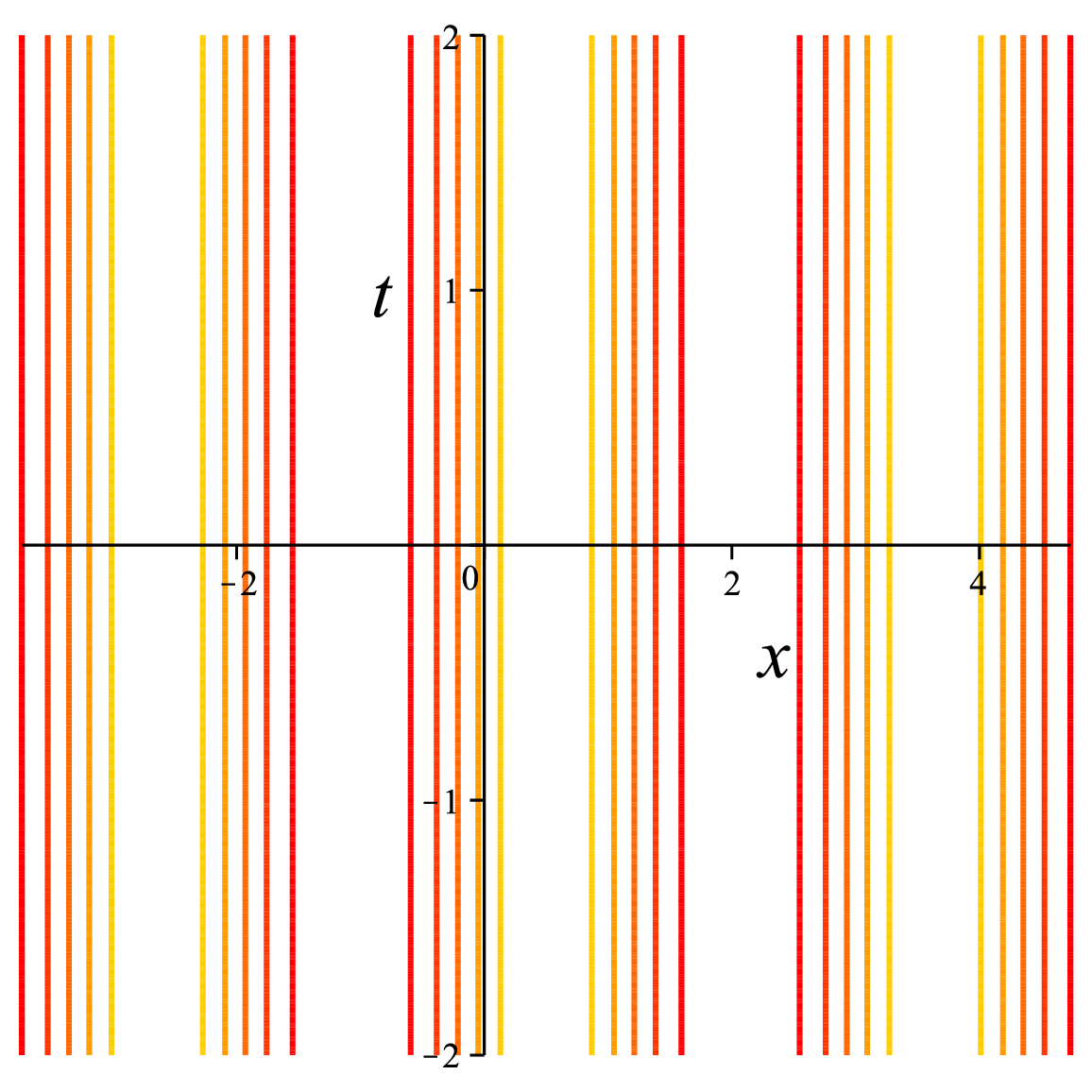}}
\caption{Periodic solution $|q|^2$ of the equation (\ref{four-ex5}) (a) 3D graph, (b) contour plot.}
\end{figure}
\end{center}
\squeezeup

\noindent \textbf{Example 13.} Consider the equation (\ref{four-ex8}) which is obtained from (\ref{4NLS-a})-(\ref{4NLS-d}) by the reduction formulas
$p(x,t)=\bar{q}(-x+x_0,t)$, $r(x,t)=\bar{q}(-x+x_0,-t+t_0)$, $s(x,t)=q(x,-t+t_0)$, and $c=\bar{c}$. Using these reductions with the solution (\ref{q-sol})-(\ref{s-sol}), we get
\begin{align}
&k_2=k_3=-\bar{k}_1,\quad k_4=k_1,\quad \omega_2=\bar{\omega}_1,\quad \omega_3=-\bar{\omega}_1,\quad \omega_4=-\omega_1,\\
&e^{\delta_2}=e^{\bar{\delta}_1+\bar{k}_1x_0},\quad e^{\delta_3}=e^{\bar{\delta_1}+\bar{k}_1x_0+\bar{\omega}_1t_0},\quad e^{\delta_4}=e^{\delta_1+\omega_1t_0}.
\end{align}
Therefore one-soliton solution of the shifted nonlocal equation (\ref{four-ex8}) is
{\small \begin{align}\label{ex13sol}
&q(x,t)=\Big(
e^{k_1x+\omega_1t+\delta_1}+\frac{\sigma(k_1+\bar k_1)e^{(2k_1-\bar k_1)x+\bar\omega_1t+2\delta_1+\bar\delta_1+\bar k_1x_0+\omega_1t_0}}
{2k_1(k_1-\bar k_1)^2}\Big)\Big/\Big(1+\frac{\sigma e^{(k_1-\bar k_1)x+\delta_1+\bar\delta_1+\bar k_1x_0}}{(k_1-\bar k_1)^2}\nonumber\\
&\times\left[
e^{(\omega_1-\bar\omega_1)t+\bar\omega_1t_0}
+
e^{-(\omega_1-\bar\omega_1)t+\omega_1t_0}
\right]+\frac{(k_1+\bar k_1)^2e^{2(k_1-\bar k_1)x+2(\delta_1+\bar\delta_1)+2\bar k_1x_0+(\omega_1+\bar\omega_1)t_0}}
{4k_1\bar k_1(k_1-\bar k_1)^4}\Big)
.
\end{align}}
Here $k_1-\bar{k}_1\neq 0$. Let $k_1=ib$, $b\neq 0$, that is $k_1+\bar{k}_1=0$, and $\delta_1=\alpha+i\beta$. Then the solution (\ref{ex13sol}) simplifies to
\begin{equation}\label{ex13solsimp}
q(x,t)=\frac{e^{ibx+\frac{b^2}{c}t+\alpha+i\beta}
}{1-\frac{\sigma}{2b^2}e^{2ibx+2\alpha-ibx_0+\frac{b^2}{c}t_0}}.
\end{equation}
This solution is singular when $\frac{\sigma}{2b^2}
e^{2ibx+2\alpha-ibx_0+\frac{b^2}{c}t_0}=1$ that is when $2b^2e^{-\left(2\alpha+\frac{b^2}{c}t_0\right)}=1$.
Note that if $\sigma=1$, the singularity occurs at $x=\frac{x_0}{2}+\frac{n\pi}{b}$, and if $\sigma=-1$, at $x=\frac{x_0}{2}+\frac{(2n+1)\pi}{2b}$ for $n\in \mathbb{Z}$.
If $2b^2e^{-\left(2\alpha+\frac{b^2}{c}t_0\right)}\neq 1$ then the solution (\ref{ex13solsimp}) is nonsingular.

Take $c=6$, $k_1=2i$, $\delta_1=1-2i$, $\sigma=1$, and
$x_0=1$, $t_0=2$. Hence the solution (\ref{ex13sol}) becomes
\begin{equation}
|q|^2=\frac{64e^{\frac{4}{3}t+2}}
{
64-16e^{\frac{10}{3}}\cos(4x-2)+e^{\frac{20}{3}}
}.
\end{equation}
The 3D and contour plot graphs of the above solution $|q|^2$ are given in the following Figure 12.
\begin{center}
\begin{figure}[h!]
\centering
\subfloat[]{\includegraphics[width=0.30\textwidth]{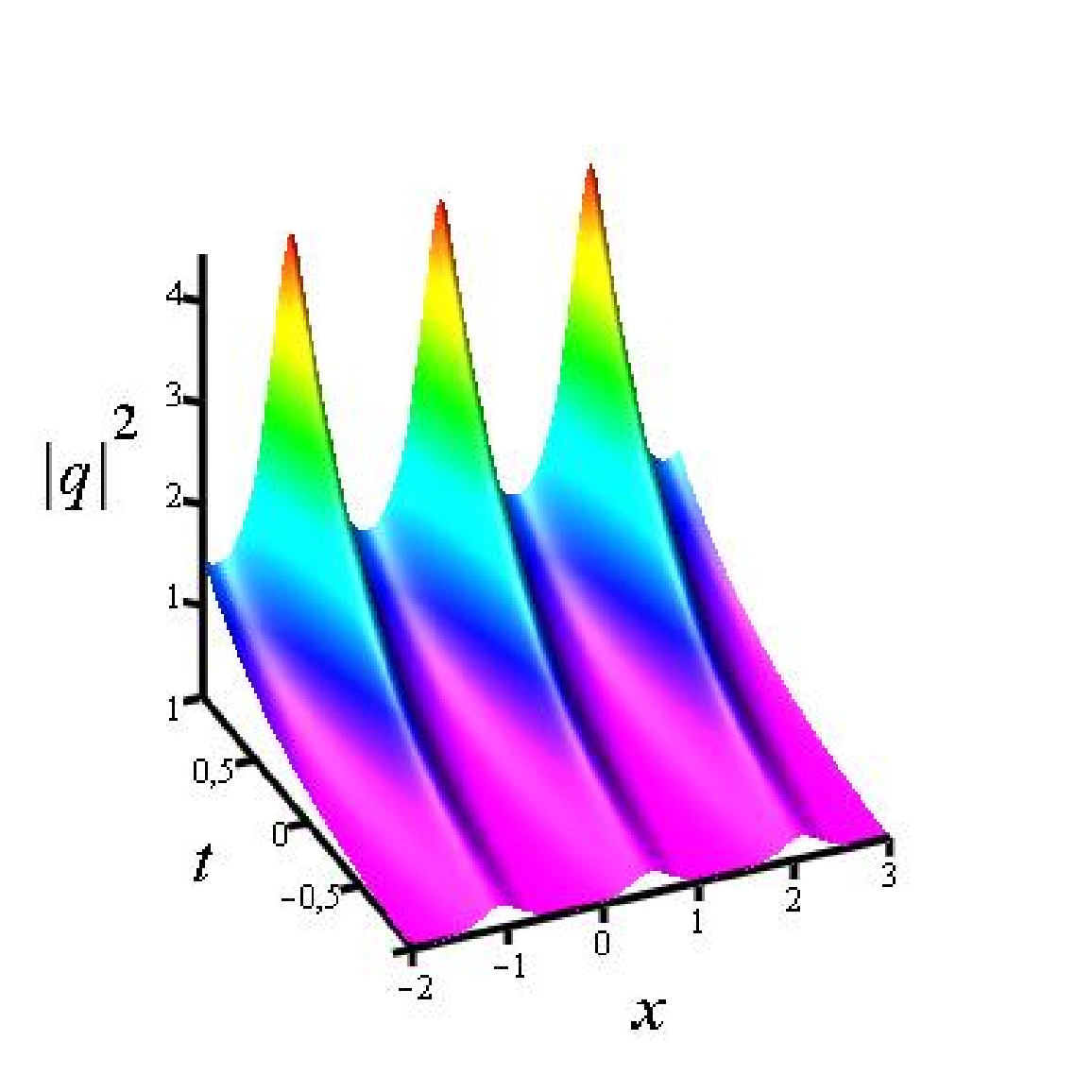}}\hspace{2cm}
\subfloat[]{\includegraphics[width=0.30\textwidth]{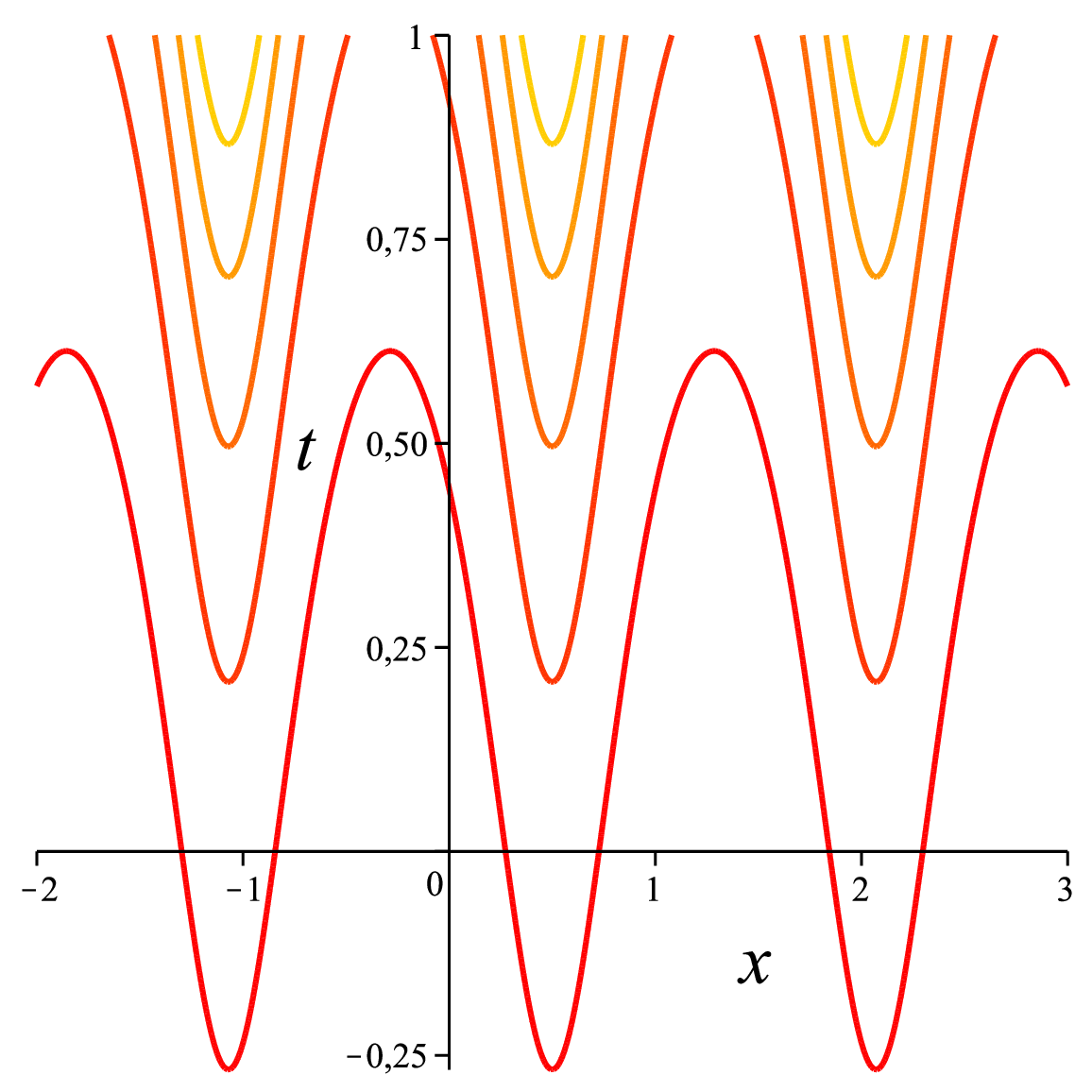}}
\caption{Periodic-type solution $|q|^2$ of the equation (\ref{four-ex8}) (a) 3D graph, (b) contour plot.}
\end{figure}
\end{center}
\squeezeup

\noindent \textbf{Example 14.} Consider the equation (\ref{four-ex10}) which is obtained from (\ref{4NLS-a})-(\ref{4NLS-d}) by the reduction formulas
$p(x,t)=\bar{q}(-x+x_0,-t+t_0)$, $r(x,t)=\bar{q}(-x+x_0,t)$, $s(x,t)=q(x,-t+t_0)$, and $c=-\bar{c}$. By using these reductions with the solution (\ref{q-sol})-(\ref{s-sol}) we obtain
\begin{align}
&k_2=k_3=-\bar{k}_1,\quad k_4=k_1,\quad \omega_2=-\bar{\omega}_1,\quad \omega_3=\bar{\omega}_1,\quad \omega_4=-\omega_1,\\
&e^{\delta_2}=e^{\bar{\delta}_1+\bar{k}_1x_0+\bar{\omega}_1t_0},\quad e^{\delta_3}=e^{\bar{\delta_1}+\bar{k}_1x_0},\quad e^{\delta_4}=e^{\delta_1+\omega_1t_0}.
\end{align}
Hence one-soliton solution of the shifted nonlocal equation (\ref{four-ex10}) is
{\small\begin{align}\label{ex14sol}
q(x,t)=\Bigg(e^{k_1x+\omega_1t+\delta_1}&+\frac{
\sigma(k_1+\bar k_1)e^{(2k_1-\bar k_1)x-\bar\omega_1t+2\delta_1+\bar\delta_1+\bar k_1x_0+(\omega_1+\bar\omega_1)t_0}
}{2k_1(k_1-\bar k_1)^2}\Bigg)\Bigg/
\nonumber\\
&\Bigg(1+\frac{\sigma e^{(k_1-\bar k_1)x+\delta_1+\bar\delta_1+\bar k_1x_0}
}
{(k_1-\bar k_1)^2}
\left[
e^{(\omega_1+\bar\omega_1)t}
+
e^{-(\omega_1+\bar\omega_1)t+(\omega_1+\bar\omega_1)t_0}
\right]
\nonumber\\&\hspace{3cm}+
\frac{
(k_1+\bar k_1)^2
e^{2(k_1-\bar k_1)x+2(\delta_1+\bar\delta_1)+2\bar k_1x_0+(\omega_1+\bar\omega_1)t_0}
}
{4k_1\bar k_1(k_1-\bar k_1)^4}\Bigg)
.
\end{align}}
Here $k_1-\bar{k}_1\neq 0$. Let $k_1=ib$, $b\neq 0$, that is $k_1+\bar{k}_1=0$, $c=i\xi$, and $\delta_1=\alpha+i\beta$. Then the solution (\ref{ex14sol}) simplifies to
\begin{equation}\label{ex14solsimp}
q=\frac{e^{ibx+\omega_1t+\alpha+i\beta}}{
1-\frac{\sigma}{2b^2}e^{2\alpha+i(2bx-bx_0)}}.
\end{equation}
 This solution is singular when its denominator $W(x)=0$, that is when $e^{2\alpha}=2b^2$ and $e^{i\theta}=\sigma$, $\theta=2bx-bx_0$. Hence if $e^{2\alpha}\neq 2b^2$ then the solution
(\ref{ex14solsimp}) is nonsingular.

Choose $c=2i$, $k_1=\frac{7}{3}i$, $\delta_1=1-2i$, $\sigma=-1$, and
$x_0=\frac{\pi}{2}$. The solution (\ref{ex14sol}) becomes
\begin{equation}
|q|^2=\frac{
9604e^2
}{
9604+81e^4
-882e^2
\left[
\sin\left(\frac{14}{3}x\right)
+
\sqrt{3}\cos\left(\frac{14}{3}x\right)
\right]
}.
\end{equation}
The 3D and contour plot graphs of the above solution $|q|^2$ are given in the following Figure 13.
\begin{center}
\begin{figure}[h!]
\centering
\subfloat[]{\includegraphics[width=0.30\textwidth]{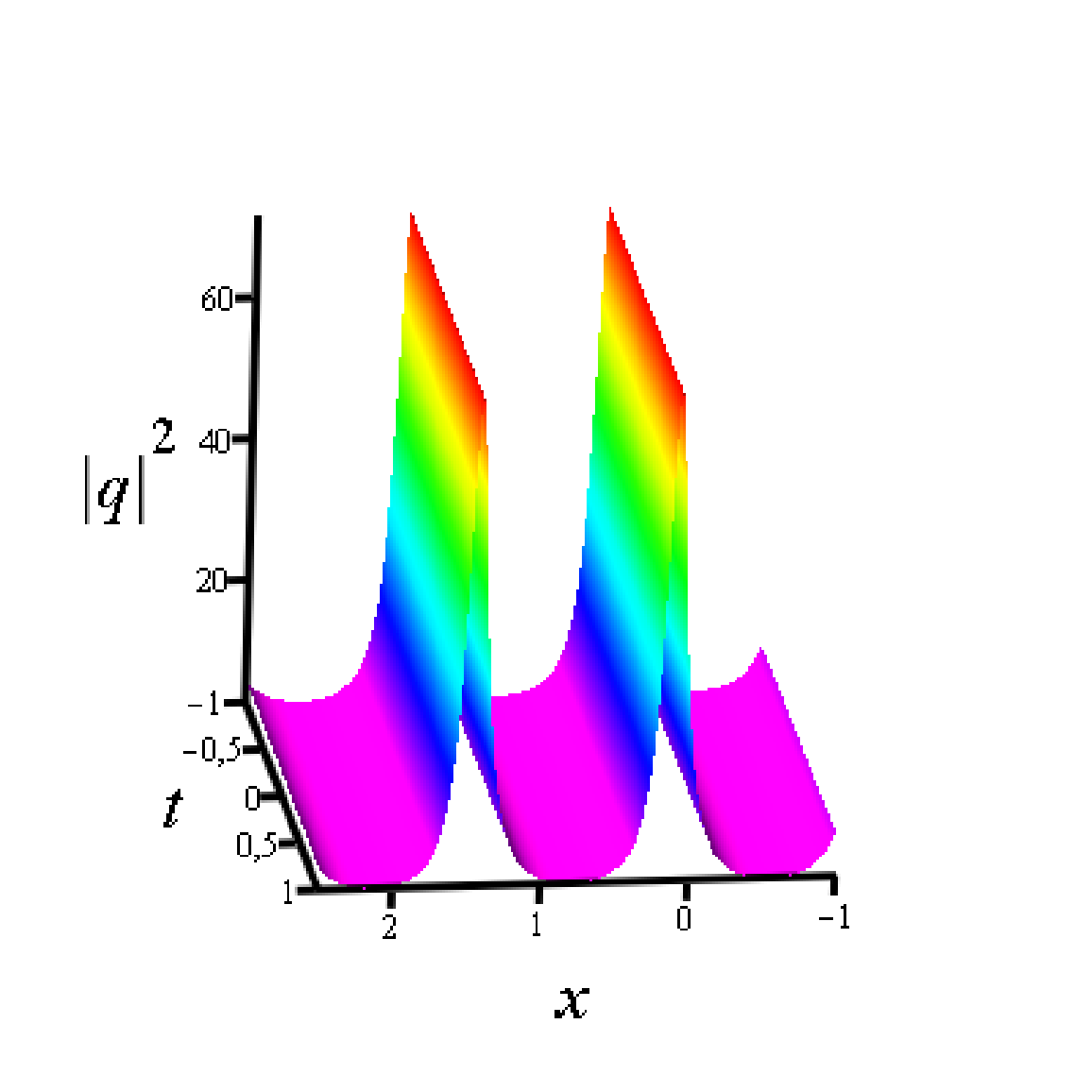}}\hspace{2cm}
\subfloat[]{\includegraphics[width=0.30\textwidth]{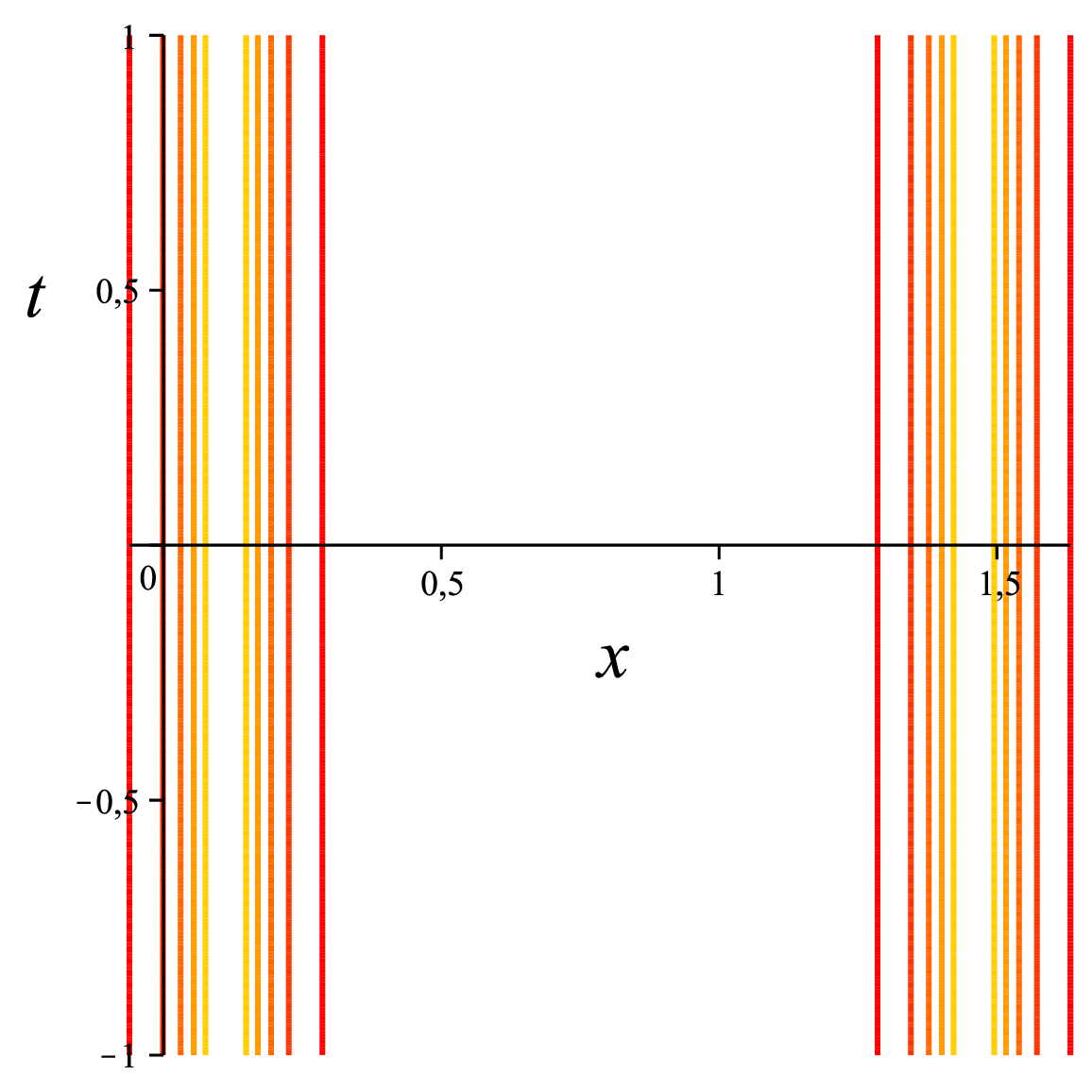}}
\caption{Periodic solution $|q|^2$ of the equation (\ref{four-ex10}) (a) 3D graph, (b) contour plot.}
\end{figure}
\end{center}
\squeezeup

\noindent \textbf{Remark.} We obtain trivial solutions for the four-place shifted nonlocal equations (\ref{four-ex1}) and (\ref{four-ex6}) by our solution method.
The constraints corresponding to the reductions leading to equations  (\ref{four-ex2}), (\ref{four-ex7}), and (\ref{four-ex9}) give $k_4=-k_1$ which makes the denominator in the general one-soliton formula vanish. Hence these cases do not yield nontrivial solutions by the present one-soliton ansatz.

\section{Conclusion}

In this work we systematically obtained the consistent shifted nonlocal two-place and four-place reductions of the four-component AKNS system. Among them, there are 13 different two-place and 10 different four-place shifted nonlocal NLS equations. The one-soliton solution of the four-component AKNS system was found through the Hirota bilinear method. By applying the shifted nonlocal reduction formulas to the obtained one-soliton solution by following our previously introduced method,  the associated constraint conditions and the explicit one-soliton solutions of the shifted nonlocal reduced equations were found out. For some cases, our method gave trivial solutions. In the case of nontrivial solutions, the singularity properties of the solutions were analyzed carefully. Through selecting proper values for the parameters, various nonsingular localized and periodic solutions have been obtained.

Our results clearly indicate that consistent shifted multi-place reductions generate many integrable nonlocal equations as well as exact solutions. Such an approach proposed here can also be generalized to other integrable multi-component AKNS systems and the integrable hierarchies. In the upcoming research, we plan to investigate different types of exact solutions like breather, lump, foldon-type solitons of the shifted multi-place nonlocal equations generated from other integrable systems.

\section{Acknowledgment}
  This work is partially supported by the Scientific
and Technological Research Council of Turkey (T\"{U}B\.{I}TAK).\\


\begin{thebibliography}{}

\bibitem{MA2019} W. X. Ma, Application of the Riemann–Hilbert approach to the
multicomponent AKNS integrable hierarchies, Nonlinear Anal.: Real World Appl. \textbf{47}, 1--17, 2019.

\bibitem{MA2002} W. X. Ma and R. Zhou, Adjoint symmetry constraints leading to binary nonlinearization, J. Nonlinear Math. Phys. \textbf{9} (1), 106--126, 2002.

\bibitem{YWL} H. X. Yang, D. L. Wang, and C. S. Li, The generalized multi-component AKNS hierarchy and N-fold Darboux transformation, Modern Phys. Lett. B \textbf{20} (25),
    1575--1589, 2006.

\bibitem{Lou} S. Y. Lou, Multi-place physics and multi-place nonlocal systems, Commun. Theor. Phys. \textbf{72}, 057001, 2020.





\bibitem{abl1} M. J. Ablowitz and Z. H. Musslimani, Integrable nonlocal nonlinear Schr\"{o}dinger equation, Phys. Rev. Lett. \textbf{110}, 064105, 2013.

\bibitem{abl2} M. J. Ablowitz and Z. H. Musslimani,  Inverse scattering transform for the integrable nonlocal nonlinear Schr\"{o}dinger equation, Nonlinearity \textbf{29}, 915--946, 2016.

    \bibitem{abl3} M. J. Ablowitz and Z. H. Musslimani, Integrable nonlocal nonlinear equations, Stud. Appl. Math. \textbf{139} (1), 7--59, 2016.

\bibitem{AbMu4} M. J. Ablowitz and Z. H. Musslimani, Integrable space-time shifted nonlocal nonlinear equations, Phys. Lett. A \textbf{409}, 127516, 2021.

\bibitem{AbMu5} M. J. Ablowitz, Z. H. Musslimani, and N. J. Ossi, Inverse scattering transform for continuous and discrete space-time shifted integrable equations, Stud. Appl. Math. \textbf{153} (4), e12764, 2024.

     \bibitem{gur1}  M. G\"{u}rses and A. Pekcan, Nonlocal nonlinear Schr\"{o}dinger equations and their soliton solutions, J. Math. Phys. \textbf{59}, 051501, 2018.

\bibitem{gur3} M. G\"{u}rses and A. Pekcan, Nonlocal nonlinear modified KdV equations and their soliton solutions,
Commun. Nonlinear Sci. Numer. Simulat. \textbf{67}, 427--448, 2019.

\bibitem{gur4} M. G\"{u}rses and A. Pekcan, Nonlocal KdV equations, Phys. Lett. A {\bf 384} (35), 126894, 2020.

\bibitem{pek2021} A. Pekcan, Local and nonlocal (2+1)-dimensional Maccari systems and their soliton solutions, Phys. Scr. {\bf 96} (3), 035217, 2021.

\bibitem{gur5}  M. G\"{u}rses and A. Pekcan, Soliton solutions of the shifted nonlocal NLS and MKdV equations, Phys. Lett. A {\bf 422}, 127793, 2022.

\bibitem{gur6} M. G\"{u}rses and A. Pekcan, $(2+1)$-dimensional local and nonlocal reductions of the
negative AKNS system: Soliton solutions, Commun. Nonlinear Sci. Numer. Simul. {\bf 71}, 161--173, 2019.

\bibitem{GPZ} M. G\"{u}rses, A. Pekcan, and K. Zheltukhin, Discrete symmetries and nonlocal reductions, Phys. Lett. A \textbf{384}, 120065, 2020.

\bibitem{Bayli}  S. Bayl{\i} and A. Pekcan, Shifted nonlocal reductions of 5-component Maccari system, Phys. Scr. \textbf{101}, 015201, 2026.

    \bibitem{Loumulti} S. Y. Lou, Multi-place physics and multi-place nonlocal systems,
Commun. Theoret. Phys. \textbf{72}, 057001, 2020.

\bibitem{SYLou1} S. Y. Lou and F. Huang, Alice-Bob physics: Coherent solutions of nonlocal KdV systems, Sci. Rep. {\bf 7}, 869, 2017.

\bibitem{SYLou2} S. Y. Lou, Alice-Bob systems, $\hat{P}-\hat{T}-\hat{C}$ symmetry invariant and symmetry breaking soliton solutions,
J. Math. Phys. \textbf{59}, 083507, 2018.


\bibitem{Mamulti} W. X. Ma, Soliton hierarchies and soliton solutions of type $(-\lambda^{\star},-\lambda)$ reduced
nonlocal nonlinear Schr\"{o}dinger equations of arbitrary even order, Partial Dif. Equ. Appl. Math. \textbf{7}, 100515, 2023.



\end{thebibliography}
\end{document}